\documentclass[12pt,a4paper]{article}

\usepackage{amssymb}
\usepackage{setspace}
\usepackage{indentfirst}

\usepackage{graphicx}

\begin{document}

\begin{titlepage}
\begin{center}

\vspace*{25mm}

\begin{spacing}{1.7}
{\LARGE\bf Polarizations of CMB and the Hubble tension}
\end{spacing}

\vspace*{25mm}

{\large
Noriaki Kitazawa
}
\vspace{10mm}

Department of Physics, Tokyo Metropolitan University,\\
Hachioji, Tokyo 192-0397, Japan\\
e-mail: noriaki.kitazawa@tmu.ac.jp

\vspace*{25mm}

\begin{abstract}
Future precision measurements of CMB polarizations
 can shed new light on the problem so called Hubble tension.
The Hubble tension comes
 from the difference of the evolutions of the Hubble parameter
 which are determined with two different distance ladders.
The standard distance ladder
 with the observation of Cepheid variables and type Ia supernovae
 gives larger values of the Hubble constant,
 and the inverse distance ladder with the observation of the baryon acoustic oscillations
 both in the CMB and in the clustering of galaxies
 gives smaller values of the Hubble constant.
These different evolutions of the Hubble parameter
 indicate different evolutions of the free electron density
 in the process of the reionization of the universe
 and different magnitudes of low-$\ell$ polarizations of the CMB,
 since these polarizations are mainly produced
 through the Thomson scattering of CMB photons off these free electrons.
We investigate the effect on
 CMB E-mode and B-mode polarizations of $\ell \leq 11$
 assuming non--trivially time--dependent equation of state of dark energy.
We find that the case of the standard distance ladder
 gives higher power of polarizations than the prediction in the $\Lambda$CDM model. 
\end{abstract}

\end{center}
\end{titlepage}

\doublespacing

\section{Introduction}
\label{sec:introduction}

The current status of the determination of the Hubble constant,
 or the present expansion rate of the universe,
 may indicate the physics beyond the $\Lambda$CDM model.
The PLANCK collaboration gives a very precise value
\begin{equation}
 H_0 = 67.4 \pm 0.5 \quad \mbox{[km/s Mpc]}
\end{equation}
 from the observation of the CMB assuming the $\Lambda$CDM model \cite{Aghanim:2018eyx}.
On the other hand,
 a direct measurement with the standard distance ladder
 using type Ia supernovae with Cepheid variables gives a precise value
\begin{equation}
 H_0 = 74.03 \pm 1.42 \quad \mbox{[km/s Mpc]}
\end{equation}
 without assuming the $\Lambda$CDM model \cite{Riess:2019cxk}.
The discrepancy of these two values is beyond $4\sigma$ in statistical significance,
 and in fact they are extremes in many other measurements and determinations.
Another direct measurement with the standard distance ladder
 using type Ia supernovae with the tip of the red giant branch method \cite{Freedman:2020dne}
 gives a value 
 $H_0 = 69.6 \pm 0.8 \pm 1.7$ [km/s Mpc]
 which does not completely agree with the value with Cepheid variables in \cite{Riess:2019cxk}.
The BOSS collaboration gives a value $H_0 = 68.6 \pm 1.1$ [km/s Mpc]
 from the signatures of Baryon Acoustic Oscillations in redshift--space galaxy surveys
 assuming the $\Lambda$CDM model \cite{Philcox:2020vvt}.
This value is obtained with the inverse distance ladder \cite{Cuesta:2014asa}
 using the determination of sound horizon scale in the CMB assuming $\Lambda$CDM model.
The value of the obtained Hubble constant is consistent with that of \cite{Aghanim:2018eyx}.

\begin{figure}[t]
\centering
\includegraphics[width=70mm]{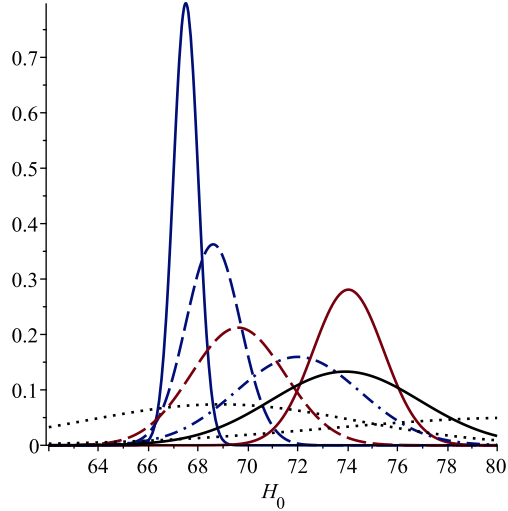}
\caption{
The current status of Hubble constant measurements.
Each measurement is shown as a normalized Gaussian distribution with a corresponding error:
the solid blue line; the PLANCK collaboration,
the dashed blue line; the BOSS collaboration,
the dash-dotted line; the SPT collaboration,
the solid red line; type Ia Supernovae with Cepheid variables,
the dashed red line; type Ia supernovae with the tip of the red giant branch method,
the solid black line; the Megamaser Cosmology Project,
and
the dotted black lines; two extreme results by the H0LiCOW collaboration.
}
\label{fig:H0}
\end{figure}

The SPT collaboration give a value $H_0 = 72.0^{+2.1}_{-2.5}$ [km/s Mpc]
 from the SPTpol survey for the CMB lensing
 combined with the signatures of Baryon Acoustic Oscillations in other galaxy surveys
 assuming the $\Lambda$CDM model \cite{Bianchini:2019vxp}.
This value is not fully consistent with that of \cite{Aghanim:2018eyx},
 though the method is essentially the same,
 looking different ranges of angular scales of the CMB perturbations.
The Megamaser Cosmology Project give a value $H_0 = 73.9 \pm 3.0$ [km/s Mpc]
 which is independent of the standard and inverse distance ladders \cite{Pesce:2020xfe},
 though still the error is not small.
The measurement using strong gravitational lensing by the H0LiCOW collaboration \cite{Wong:2019kwg} 
 is also independent of the standard and inverse distance ladders,
 and they give a value $H_0 = 73.3^{+1.7}_{-1.8}$ [km/s Mpc],
 but it is the statistical combination of six values of measurements on different quasars
 distributing a wide range
 between $H_0 = 81.1^{+8.0}_{-7.1}$ and $H_0 = 68.9^{+5.4}_{-5.1}$ [km/s Mpc],
 and we need to wait more measurements.
These measurements are summarized in Fig.\ref{fig:H0},
 and our intension here does not give a complete summary,
 but gives a simple sketch of current status.
This situation so called Hubble tension
 could merely happen as statistical fluctuations or by some systematic errors,
 but it might suggest something worth to investigate.

Since there is a tendency that
 the lower redshift measurements give larger values of the Hubble constant,
 we may consider only the lower redshift physics
 not to disturb the success of the $\Lambda$CDM model for higher redshift physics,
 as it has been extensively discussed in \cite{Raveri:2019mxg}.
On the other hand,
 since there could be some internal inconsistencies
 in the measurements between high and low multipoles
 and also in the measurements of weak lensing effects
 by the PLANCK collaboration
 \cite{Aghanim:2018eyx,DiValentino:2019qzk,Handley:2019tkm,Efstathiou:2020wem,DiValentino:2020hov,
 Vagnozzi:2020zrh,Vagnozzi:2020dfn},
 it may be required more extensive revision of the $\Lambda$CDM model
 (see \cite{Knox:2019rjx} for a review).
In this work we consider the former possibility and
 investigate the chance to see some evidences of physics beyond the $\Lambda$CDM model
 in future precise measurements of the low-$\ell$ polarizations of the CMB
 by LiteBIRD, for example.

We may consider that
 the dark energy is not the cosmological constant,
 but something else which can be captured by considering time--dependent
 equation of state $p=w\rho$ like
\begin{equation}
 w = w_0 + w_a (1-a) = w_0 + w_a \frac{z}{1+z},
\end{equation}
 where and throughout this paper
 the scale factor is normalized as $a=1$ at present.
This model is called Chevallier-Polarski-Linder (CPL) model
 \cite{Chevallier:2000qy,Linder:2002et}.
The cosmological constant is represented simply by $w_0=-1$ and $w_a=0$.
The value of $w_0$ represents the present value of $w$
 and $w_a$ describes a time dependence of the equation of state.
The evolution of the Hubble parameter is given as
\begin{equation}
 H(z) = H_0 \sqrt{\Omega_m (1+z)^3 + \Omega_{\rm DE} (1+z)^{3(1+w_0+w_a)} e^{-3w_a\frac{z}{1+z}}},
\label{Hubble-CPL}
\end{equation}
 where we set $\Omega_m=0.3$ and $\Omega_{\rm DE}=1-\Omega_m=0.7$
 so that it reduces to that of the $\Lambda$CDM model in case of $w_0=-1$ and $w_a=0$.

We also consider a simple Taylor expansion
\begin{equation}
 w = w_0 + w_1 a + \frac{1}{2} w_2 a^2 = w_0 + w_1 \frac{1}{1+z} + \frac{1}{2} w_2 \frac{1}{(1+z)^2}
\end{equation}
 with $w_0=-1$.
The intention is to capture small deviation from the case of the cosmological constant
 by two parameters, $w_1$ and $w_2$.
The amount of the deviation can be larger at recent $a \sim 1$ than at far past $a \sim 0$.
The evolution of the Hubble parameter is described as
\begin{equation}
 H(z) = H_0 \sqrt{\Omega_m (1+z)^3
                   + \Omega_{\rm DE} (1+z)^{3(1+w_0)}
                     e^{3w_1(1-\frac{1}{1+z})+\frac{3}{4}w_2(1-\frac{1}{(1+z)^2})}}
\label{Hubble-Taylor}
\end{equation}
 with $w_0=-1$, $\Omega_m=0.3$ and $\Omega_{\rm DE}=1-\Omega_m=0.7$.

We perform two simple fits of $H_0$, $w_0$ and $w_a$ with two sets of data:
 one is obtained with the standard distance ladder
 and the other is obtained with the inverse distance ladder.
The data set of the standard distance ladder consists
 the values of Hubble parameters in \cite{Riess:2019cxk} and \cite{Freedman:2020dne}
 (they use the same set of type Ia supernovae in the Large Magellanic Cloud
  with different distance calibration methods)
 and five determinations of the Hubble parameter at different redshifts
 using Pantheon + MCT type Ia supernovae in \cite{Riess:2017lxs}.
The data set of the inverse distance ladder consists
 four values of the Hubble parameter by the BOSS collaboration at different redshifts in
 \cite{Alam:2016hwk,Gil-Marin:2018cgo,Percival:2018twa,Agathe:2019vsu}
 and the PLANCK result of the Hubble constant in \cite{Aghanim:2018eyx}
 which we translate to the Hubble parameter at $z=1000$ assuming the $\Lambda$CDM model.

\begin{figure}[t]
\centering
\includegraphics[width=50mm]{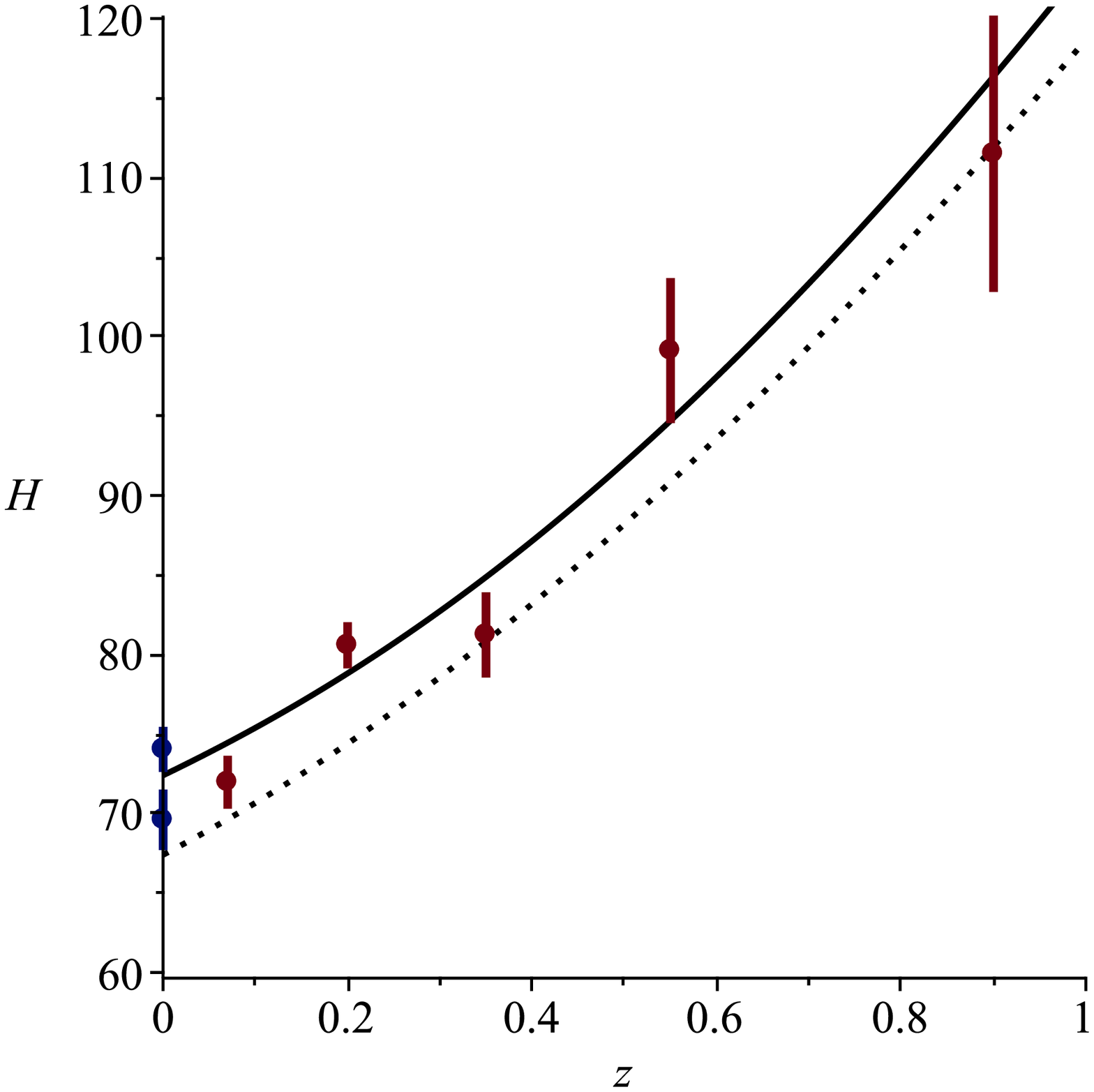}
\hspace{1cm}
\includegraphics[width=50mm]{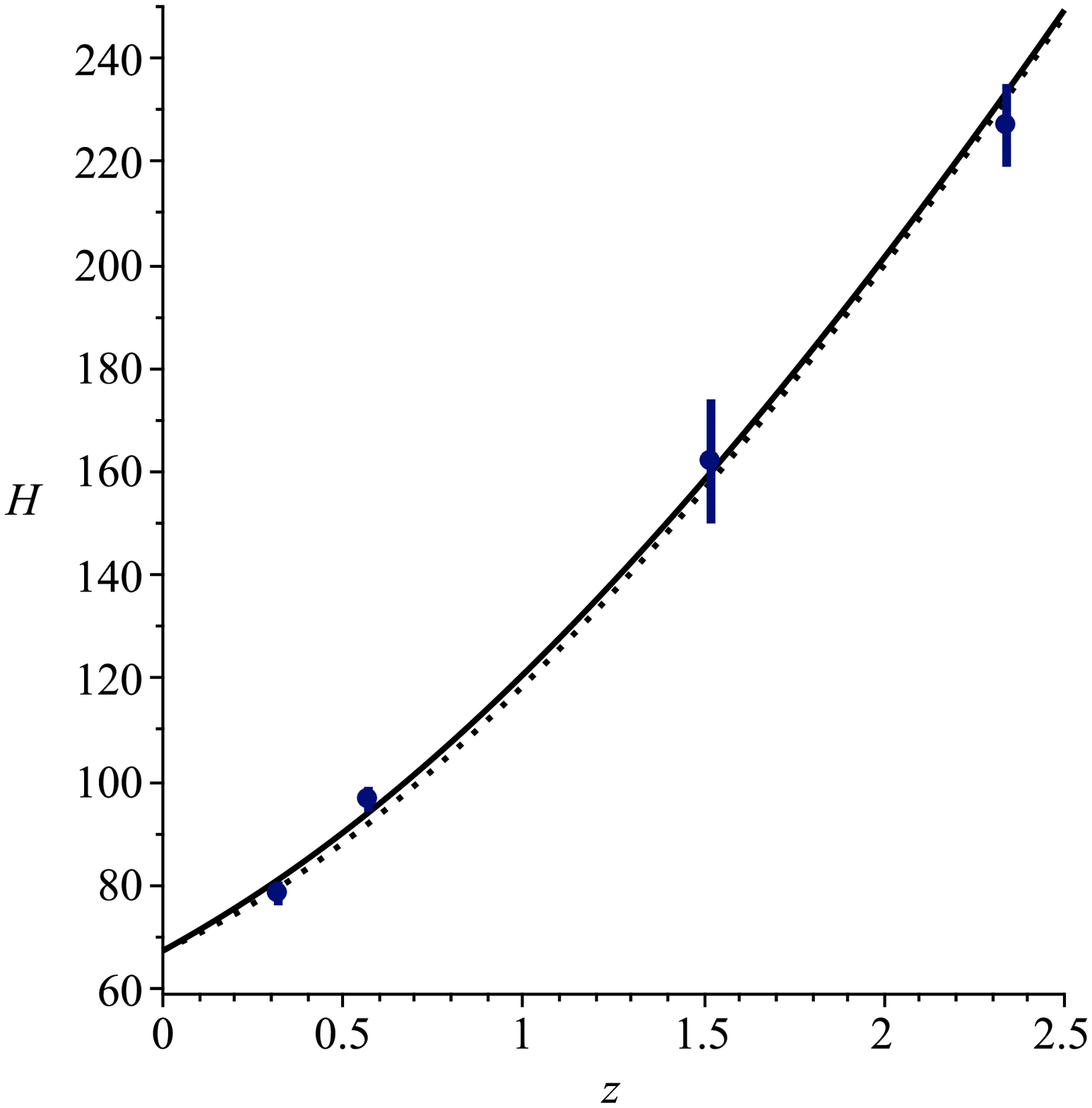}
\caption{
Fits of parameters in the CPL model.
The solid lines indicate the results of fits
 and dotted lines indicate the concordance $\Lambda$CDM expectation by the PLANCK collaboration.
Left: the fit with the data set of the standard distance ladder
 which gives $H_0=72.4$ [km/s Mpc], $w_0=-1.06$ and $w_a=-0.34$.
Right: the fit with the data set of the inverse distance ladder
 which gives $H_0=67.3$ [km/s Mpc], $w_0=-0.88$ and $w_a=-0.15$.
}
\label{fig:fit-CPL}
\end{figure}
\begin{figure}[t]
\centering
\includegraphics[width=50mm]{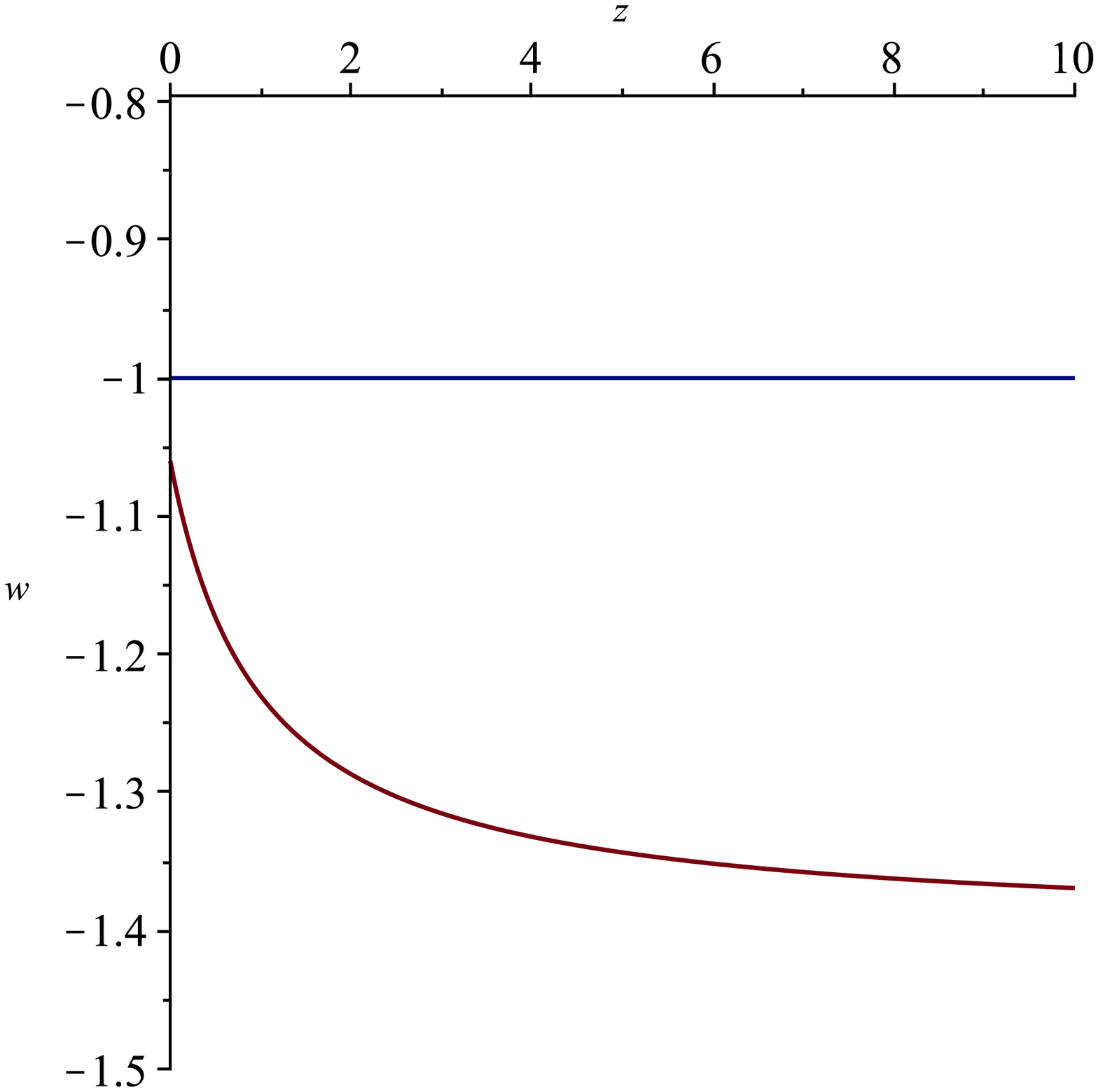}
\hspace{1cm}
\includegraphics[width=50mm]{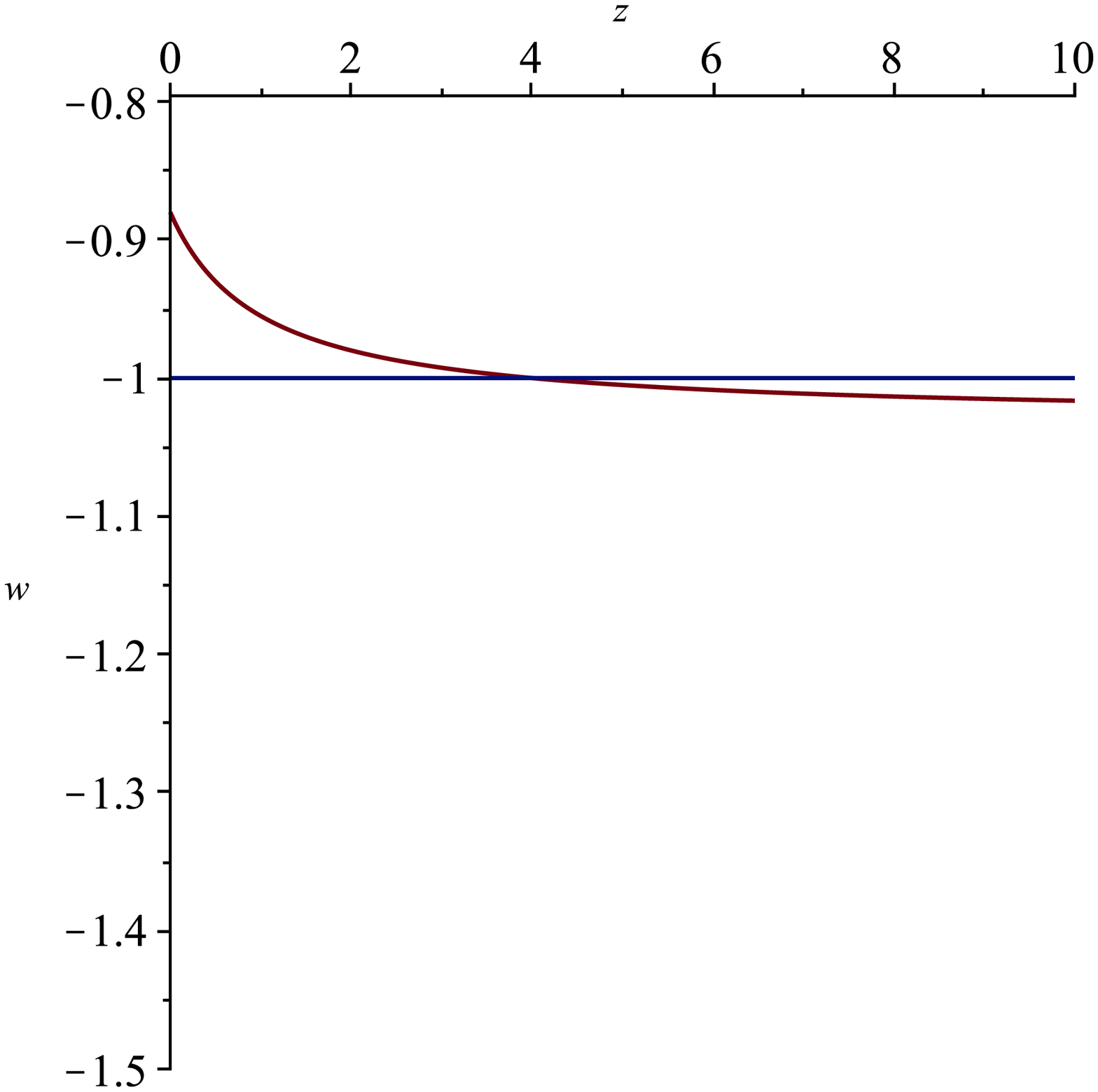}
\caption{
Redshift--dependences of $w$ in the CPL model.
Left: the case of the standard distance ladder ($w_0=-1.06$ and $w_a=-0.34$).
Right: the case of the inverse distance ladder ($w_0=-0.88$ and $w_a=-0.15$).
}
\label{fig:w-CPL}
\end{figure}
Fig.\ref{fig:fit-CPL} shows the results of our simple least $\chi^2$ fits for the CPL models.
The obtained sets of the values of $w_0$ and $w_a$
 for the standard distance ladder and the inverse distance ladder
 are consistent with the results of rigorous fit results
 in \cite{Riess:2017lxs} and \cite{Aghanim:2018eyx}, respectively.
The corresponding redshift--dependences of $w$ are shown in Fig.\ref{fig:w-CPL}.
We see that
 the deviation from the $\Lambda$CDM model
 is larger for the case of the standard distance ladder than the case of inverse distance ladder.
In other words we could say that
 the case of the inverse distance ladder essentially suggests the $\Lambda$CDM model
  and the case of the standard distance ladder suggests some new physics at low-$z$
  on the other hand.
 
\begin{figure}[t]
\centering
\includegraphics[width=50mm]{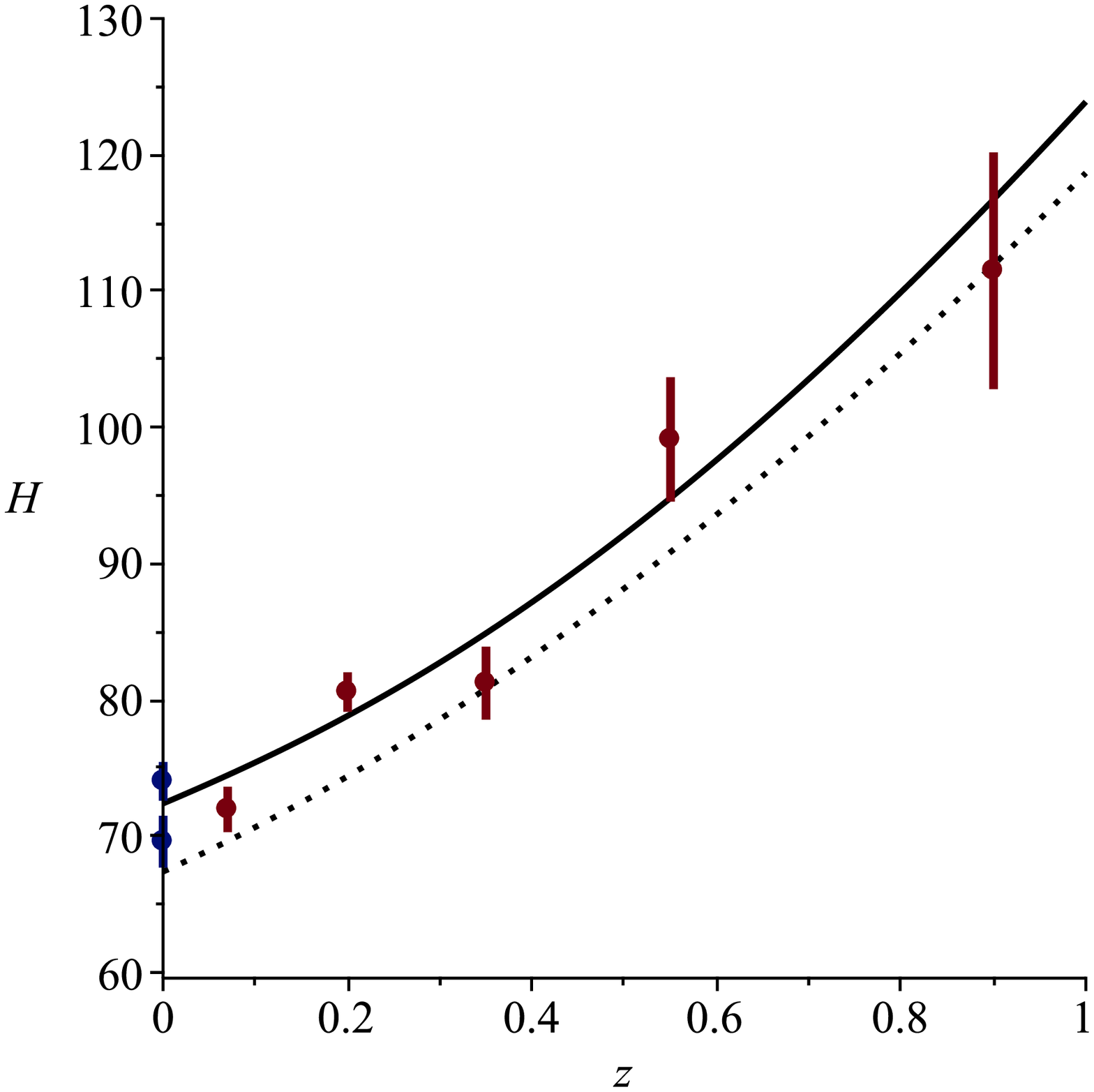}
\hspace{1cm}
\includegraphics[width=50mm]{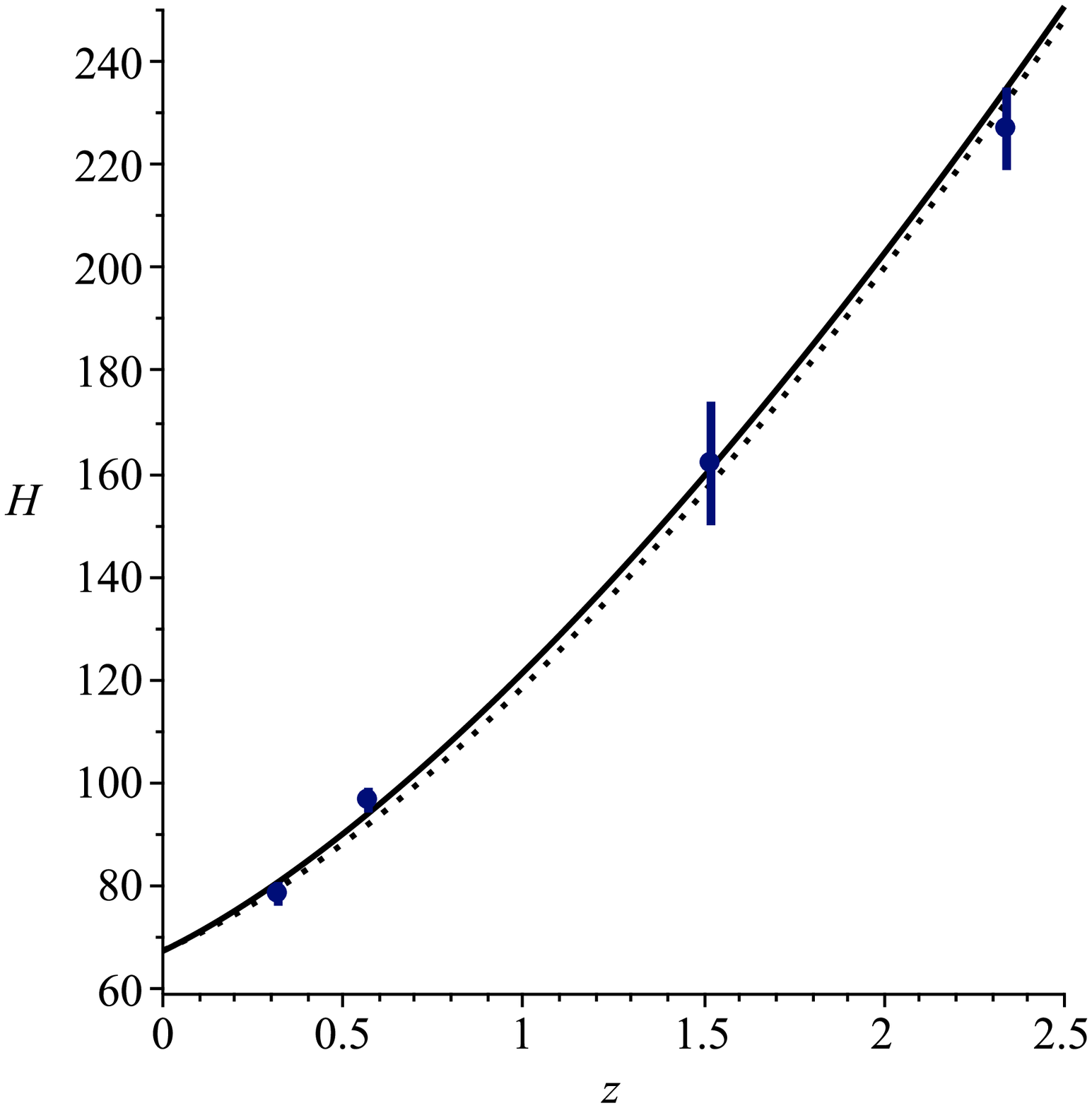}
\caption{
Fits of parameters of the model of Taylor expansion.
The solid lines indicate the results of fits
 and dotted lines indicate the concordance $\Lambda$CDM expectation by the PLANCK collaboration.
Left: the fit with the data set of the standard distance ladder
 which gives $H_0=72.4$ [km/s Mpc], $w_1=-0.60$ and $w_2=1.1$.
Right: the fit with the data set of the inverse distance ladder
 which gives $H_0=67.3$ [km/s Mpc], $w_1=0.4$ and $w_2=-0.7$.
}
\label{fig:fit-Taylor}
\end{figure}
\begin{figure}[t]
\centering
\includegraphics[width=50mm]{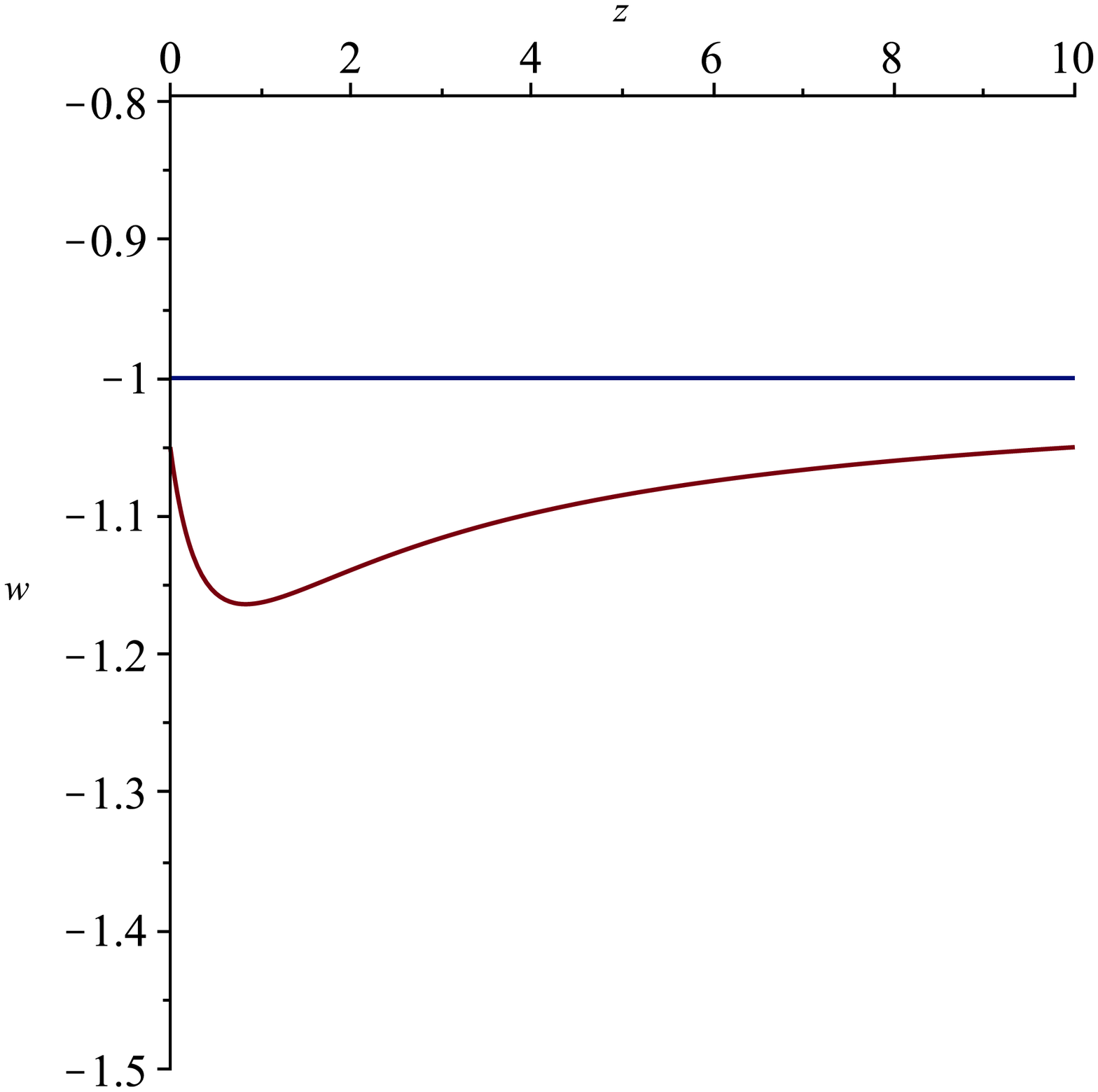}
\hspace{1cm}
\includegraphics[width=50mm]{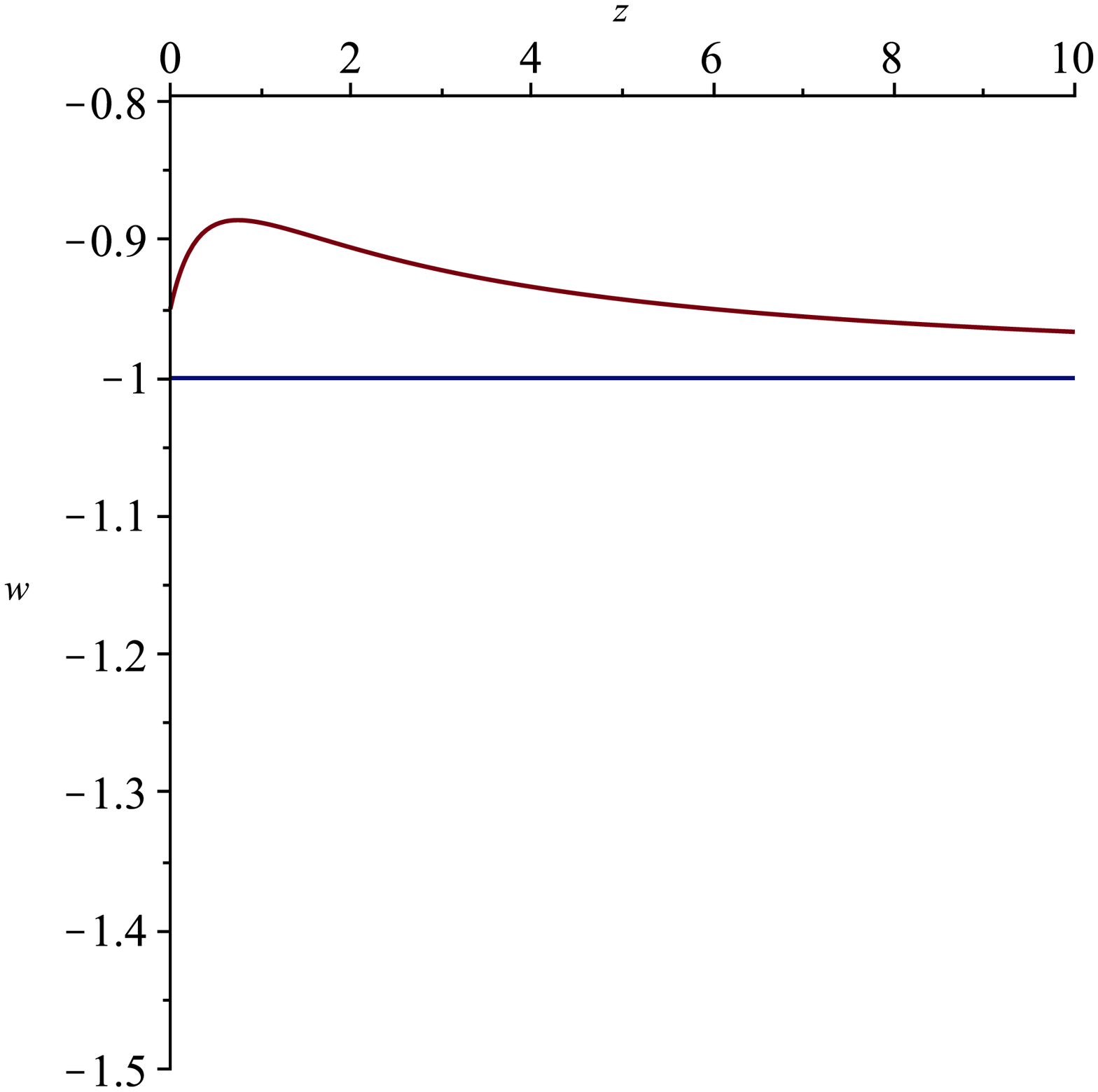}
\caption{
Redshift--dependences of $w$ in the model of Taylor expansion.
Left: the case of the standard distance ladder ($w_1=-0.60$ and $w_2=1.1$).
Right: the case of the inverse distance ladder ($w_1=0.4$ and $w_2=-0.7$).
}
\label{fig:w-Taylor}
\end{figure}
Figs.\ref{fig:fit-Taylor} and \ref{fig:w-Taylor}
 show the same for the model of Taylor expansion of $w$.
Though the shapes of $H(z)$ are almost the same of those in the CPL model,
 the redshift--dependence of $w$ are different.
In the model of Taylor expansion $w$ quickly goes to $-1$
 at larger values of redshift, but it is not the case in the CPL model.
Therefore,
 in view of modifying only low--redshift physics the model of Taylor expansion may be better.
It is also noteworthy that
 $w < -1$ for the case of the standard distance ladder
 and $w \gtrsim -1$ for the case of the inverse distance ladder.
Here, we note that these behaviors of $w$ at larger redshift values
 are extrapolations depending on models
 which are constrained only by the data at low redshift values $z \lesssim 1$.
Since the universe is dark energy dominant at $z \lesssim 1$, however,
 investigating the nature of dark energy with low--redshift data
 is reasonable at present status without precise information of the universe at larger redshift values.

In the next section
 we introduce the semi--analytic method
 to calculate angular power spectra $C^{EE}_\ell$ and $C^{BB}_\ell$ for $\ell \leq 11$
 which has been developed in
 \cite{Polnarev:1985,Basko:2080,Harari:1993nb,Zaldarriaga:1995gi,Ng:1993fx,Ng:1994sv,Kamionkowski:1996ks}.
We utilize the approximation so called long wave length limit, or tight coupling limit
 \cite{Zaldarriaga:1995gi,Pritchard:2004qp,Zhang:2005nv,Kitazawa:2019fzc}.
Though this approximation is rather drastic and rigorous quantitative precision is sacrificed,
 the resultant semi--analytic method clearly highlights the underlying physics.
In section \ref{sec:results}
 we apply the method to the CPL models with the standard distance ladder and the inverse distance ladder,
 as well as for the models of Taylor expansion.
We see that without model dependence (either the CPL model or the model of Taylor expansion)
 both $D^{EE}_\ell \equiv \ell(\ell+1)C^{EE}_\ell/2\pi$ and $D^{BB}_\ell \equiv \ell(\ell+1)C^{BB}_\ell/2\pi$
 are somewhat enhanced with respect to the $\Lambda$CDM prediction
 at $8 < \ell < 11$ in case of the standard distance ladder,
 and slightly suppressed with respect to the $\Lambda$CDM prediction
 at $8 < \ell < 11$ in case of the inverse distance ladder.
This indicate that
 future precise polarization measurements can give new information about the Hubble tension
 and physics beyond the $\Lambda$CDM model.
In the last section we summarize our results and conclude.
The readers who are not interested in the theoretical formalism
 can skip the next section and proceed to section \ref{sec:results}.

\section{The semi--analytic method}
\label{sec:method}

The reionization of the universe,
 which is expected to happen in the redshift period of $6 \lesssim z \lesssim 10$,
 gives free electrons by which CMB photons get Thomson scattering. 
In the presence of CMB perturbations
 such a Thomson scattering produces polarizations of CMB perturbations:
 the scalar perturbation produces E-mode polarization
 and the tensor perturbation produces both E-mode and B-mode polarizations.
Since the tensor perturbation, which has not yet been observed,
 is much smaller than the scalar perturbations,
 we concentrate on E-mode polarization by the scalar perturbation only
 neglecting the contribution of the tensor perturbation.
In this work we set the tensor-to-scalar ratio as $r_{\rm TS}=0.03$.

We first introduce a semi--analytic method
 to calculate the angular power spectrum of E-mode polarization $D^{EE}_\ell$ due to scalar perturbations.
We take the synchronous gauge:
\begin{equation}
 ds^2 = a(\eta)^2 \left[ d\eta^2 - (\delta_{ij} + h_{ij}) dx^i dx^j \right],
\label{metric}
\end{equation}
 where $\eta$ is the conformal time and $h_{ij}$ describes the perturbations around the background.
The array of radiation strength is introduced as
\begin{equation}
 {\small
  \left(
  \begin{array}{c}
   I_\theta \\
   I_\varphi \\
   U
  \end{array}
 \right)
 }
 = \frac{h\nu^3}{c^2} \tilde{f},
\qquad
 \tilde{f} = f_0(T_0,\nu)
  {\small \left[ \left( \begin{array}{c} 1 \\ 1 \\ 0 \end{array} \right) + \tilde{f}_1 \right]},
\end{equation}
 where $I_\theta$ and $I_\varphi$ are intensities of the CMB radiation
 corresponding to two independent directions of polarizations
 and $U$ is one of the Stokes parameters \cite{chandrasekhar:1960}.
Other stokes parameters are described as $Q=I_\theta - I_\varphi$ and $V=0$
 (Thomson scattering does not produce circular polarizations).
Since the CMB radiation
 is almost the blackbody radiation of temperature $T_0 \simeq 2.7$[K] with small perturbations,
 we can write the array in this way using $f_0(T_0,\nu) = 1/(\exp(h\nu/k_B T_0)-1)$
 and an array of small perturbations $\tilde{f}_1$.
If we assume that
 the perturbation of intensity $\delta I$ around the blackbody background $I_0$
 is described as the temperature perturbation of the blackbody radiation intensity,
\begin{equation}
 I_0 + \delta I = \frac{h\nu^3}{c^2} f_0(T_0+\delta T,\nu)
  \simeq \frac{h\nu^3}{c^2} f_0(T_0,\nu) \left( 1 - \gamma \frac{\delta T}{T_0} \right),
\label{perturbation-assumption}
\end{equation}
 where
\begin{equation}
 \gamma \equiv \frac{\nu}{f_0} \frac{\partial f_0}{\partial \nu},
\end{equation}
 we may write the array of perturbations as $\tilde{f}_1 = \gamma \delta \tilde{f}$.
Note that the perturbation $\delta \tilde{f}$ does not depend on the frequency of radiation,
 though $\gamma$ is a function of frequency.

The Boltzmann equation (or the equation of radiation transfer divided by $h\nu^3/c^2$) is
\begin{equation}
 \frac{\partial \tilde{f}}{\partial \eta}
 - \hat{n}_i \frac{\partial \tilde{f}}{\partial x^i}
 - \frac{1}{2} \frac{\partial h_{ij}}{\partial \eta} \hat{n}_i \hat{n}_j
   \nu \frac{\partial \tilde{f}}{\partial \nu}
 = - g(\eta) \left( \tilde{f} - \tilde{J} \right),
\label{Boltzmann-eq}
\end{equation}
 where $\hat{n} = (\sin\theta \cos\varphi, \sin\theta\sin\varphi, \cos\theta)$
 is the unit vector pointing toward the sky in the direction of the photon to be observed, and 
\begin{equation}
 \tilde{J}
  = \frac{1}{4\pi} \int_{-1}^{1} d\mu' \int_0^{2\pi} d\varphi'
    \tilde{P}(\mu,\varphi,\mu',\varphi') \tilde{f}(\mu,\mu')
\end{equation}
 which describes the effect of Thomson scattering, where $\mu \equiv \cos\theta$.
The function $g(\eta) \equiv \sigma_T n_e(\eta) a(\eta)$,
 where $\sigma_T$ is the Thomson scattering cross section
 and $n_e(\eta)$ is the density of free electrons, which are produced during reionization,
 describes the efficiency of the Thomson scattering.
The third term of the left hand side of eq.(\ref{Boltzmann-eq}) describes Sachs-Wolfe effect
 which is proportional to $\gamma$ in the first order of perturbation
 (the first order in $\tilde{f}_1$ and $h_{ij}$).

Consider only the scalar perturbation of the background metric.
The gauge--invariant primordial scalar perturbation is produced in the period of inflation as
\begin{equation}
 {\cal R}^0 (x) = \int \frac{d^3k}{(2\pi)^3}
  \left(
   \hat{\alpha}_{\bf k} {\cal R}^0_k e^{-i{\bf k} \cdot {\bf x}}
   + \hat{\alpha}_{\bf k}^\dag ({\cal R}^0_k)^* e^{i{\bf k} \cdot {\bf x}}
  \right),
\end{equation} 
 where $\hat{\alpha}_{\bf k}$ is the stochastic variable which satisfies
 $\langle \hat{\alpha}_{\bf k} \hat{\alpha}_{\bf k'}^\dag \rangle
  = (2\pi)^3 \delta^3({\bf k} - {\bf k'})$.
The primordial power spectrum $P_{\cal R}(k)$ is defined as
\begin{equation}
 \langle {\cal R}^0(x) {\cal R}^0(x) \rangle
  = \int_0^\infty \frac{dk}{k} \frac{k^3}{2\pi^2} \vert {\cal R}^0_k \vert^2
 \equiv \int_0^\infty \frac{dk}{k} P_{\cal R}(k)
\end{equation}
 and its relevant parameterization is 
\begin{equation}
 P_{\cal R}(k) = A_S \left( \frac{k}{k_{\rm pivot}} \right)^{n_s-1}
\end{equation}
 with $A_S \simeq 2.1 \times 10^{-9}$ and $n_s \simeq 0.965$ at $k_{\rm pivot} = 0.05$ [Mpc${}^{-1}$]
 which are determined by the PLANCK collaboration \cite{Aghanim:2018eyx}.
By solving evolution equations of scalar perturbations in synchronous gauge we obtain
\begin{equation}
 \frac{\partial h_{ij}}{\partial \eta}
  = \frac{2}{15} \frac{\partial}{\partial x^i} \frac{\partial}{\partial x^j} {\cal R}^0 (x)
\end{equation}
 which coincides with that given in \cite{Harari:1993nb} up to irrelevant overall sign.

The intensity perturbations are also expanded with the same stochastic variable as
\begin{equation}
 \tilde{f}_1(x) = \int \frac{d^3k}{(2\pi)^3}
  \left(
   \hat{\alpha}_{\bf k} \tilde{f}_{1k} e^{-i{\bf k} \cdot {\bf x}}
   + \hat{\alpha}_{\bf k}^\dag (\tilde{f}_{1k})^* e^{i{\bf k} \cdot {\bf x}}
  \right),
\end{equation}
 and if we consider one component of scalar perturbations which propagates $z$ direction:
 ${\cal R}^0_k e^{-ikz}$, the Boltzmann equation becomes the equation
 for the corresponding component $\tilde{f}_{1k} e^{-ikz}$.
Since there is a symmetry of rotation around $z$-axis in this case, $U=0$,
 namely no B-mode polarizations.
The matrix of the Thomson scattering in this case is \cite{chandrasekhar:1960}
\begin{equation}
 P(\mu,\mu')
  = \frac{3}{4} \left(
                 \begin{array}{cc}
                 2(1-\mu^2)(1-\mu'^2)+\mu^2 \mu'^2 & \mu^2 \\
                 \mu'^2 & 1
                 \end{array}
                \right),
\end{equation}
 and the Boltzmann equation for the first order perturbations becomes
\begin{equation}
 \left(
  \frac{\partial}{\partial\eta} + i k \mu + g(\eta)
 \right) \tilde{f}_{1k}(\eta,\mu,\nu)
 + \frac{\gamma}{15} \eta (k\mu)^2 {\cal R}^0_k
   {\scriptsize \left( \begin{array}{c} 1 \\ 1 \end{array} \right)}
 = g(\eta) \frac{1}{2} \int_{-1}^1 d\mu' P(\mu,\mu') \tilde{f}_{1k}(\eta,\mu',\nu).
\end{equation}
Note that this equation depends on the observing frequency of the CMB,
 and the resultant amount of the polarizations depends on the frequency,
 because Sachs-Wolfe effect depends on the frequency.
If we follow the assumption of eq.(\ref{perturbation-assumption})
 and replace $\tilde{f}_1$ by $\gamma \delta {\tilde f}$,
 we have the Boltzmann equation which is independent from the frequency:
\begin{equation}
 \left(
  \frac{\partial}{\partial\eta} + i k \mu + g(\eta)
 \right) \delta \tilde{f}_k(\eta,\mu)
 + \frac{1}{15} \eta (k\mu)^2 {\cal R}^0_k
   {\scriptsize \left( \begin{array}{c} 1 \\ 1 \end{array} \right)}
 = g(\eta) \frac{1}{2} \int_{-1}^1 d\mu' P(\mu,\mu') \delta \tilde{f}_k(\eta,\mu).
\end{equation}

Now decompose the array of the perturbation as
\begin{equation}
 \delta \tilde{f}_k
  = \alpha_k
    {\small
    \left(
     \begin{array}{c}
      1 \\ 1
     \end{array}
    \right)
    }
  + \beta_k
    {\small
    \left(
     \begin{array}{c}
      1 \\ -1
     \end{array}
    \right)
    },
\end{equation}
 where $\alpha_k$ and $\beta_k$ describe
 the perturbations of the total intensity and Stokes parameter $Q$, respectively.
We obtain the equations
\begin{equation}
 \left(
  \frac{\partial}{\partial\eta} + i k \mu + g(\eta)
 \right) \alpha_k(\eta,\mu)
 = - \frac{{\cal R}^0_k}{15} \eta (k\mu)^2
   + g(\eta) \left\{ \alpha_{k,0} - \left( \mu^2 - \frac{1}{3} \right) G_k(\eta) \right\},
\label{eq-for-alpha}
\end{equation}
\begin{equation}
 \left(
  \frac{\partial}{\partial\eta} + i k \mu + g(\eta)
 \right) \beta_k(\eta,\mu)
 = g(\eta) \left( 1- \mu^2 \right) G_k(\eta),
\label{eq-for-beta}
\end{equation}
 where
\begin{equation}
 G_k(\eta) \equiv \frac{3}{4} \left( \beta_{k,0} - \alpha_{k,2} - \beta_{k,2} \right)
\end{equation}
 is the so called source function and
\begin{equation}
 \alpha_k(\eta,\mu) = \sum_{\ell=0}^\infty (2\ell+1) \alpha_{k,\ell}(\eta) P_\ell(\mu),
\qquad
 \beta_k(\eta,\mu) = \sum_{\ell=0}^\infty (2\ell+1) \beta_{k,\ell}(\eta) P_\ell(\mu).
\end{equation}
These eqs.(\ref{eq-for-alpha}) and (\ref{eq-for-beta}) can be described in a different form as follows.
The equations or $\alpha_{k,\ell}(\eta)$
\begin{eqnarray}
 \dot{\alpha}_{k,0} 
 &=& - ik \alpha_{k,1} - \frac{{\cal R}^0_k}{45} \eta k^2,
\\
 \dot{\alpha}_{k,1}
 &=& - ik \left( \frac{1}{3} \alpha_{k,0} + \frac{2}{3} \alpha_{k,2} \right) - g \alpha_{k,1},
\\
 \dot{\alpha}_{k,2}
 &=& - ik \left( \frac{2}{5} \alpha_{k,1} + \frac{3}{5} \alpha_{k,3} \right)
     - \frac{2}{5}\frac{{\cal R}^0_k}{45} \eta k^2
     - g \frac{1}{10} \left( 9 \alpha_{k,2} + \beta_{k,0} - \beta_{k,2} \right),
\\
 \dot{\alpha}_{k,\ell}
 &=& - ik \left( \frac{\ell}{2\ell+1} \alpha_{k,\ell-1} + \frac{\ell+1}{2\ell+1} \alpha_{k,\ell+1} \right)
     - g \alpha_{k,\ell},
 \quad \ell \geq 3,
\end{eqnarray}
 where dots indicate derivatives by $\eta$, and the equations for $\beta_{k,\ell}(\eta)$
\begin{eqnarray}
 \dot{\beta}_{k,0} 
 &=& - ik \beta_{k,1} - g \frac{1}{2} \left( \beta_{k,0} + \alpha_{k,2} + \beta_{k,2} \right),
\\
 \dot{\beta}_{k,1}
 &=& - ik \left( \frac{1}{3} \beta_{k,0} + \frac{2}{3} \beta_{k,2} \right) - g \beta_{k,1},
\\
 \dot{\beta}_{k,2}
 &=& - ik \left( \frac{2}{5} \beta_{k,1} + \frac{3}{5} \beta_{k,3} \right)
     - g \frac{1}{10} \left( 9 \beta_{k,2} + \beta_{k,0} - \alpha_{k,2} \right),
\\
 \dot{\beta}_{k,\ell}
 &=& - ik \left( \frac{\ell}{2\ell+1} \beta_{k,\ell-1} + \frac{\ell+1}{2\ell+1} \beta_{k,\ell+1} \right)
     - g \beta_{k,\ell},
 \quad \ell \geq 3.
\end{eqnarray}
The solution of eq.(\ref{eq-for-beta}) can be formally written as
\begin{equation}
 \beta_k
  = \int_0^\eta d\eta' e^{-\kappa(\eta,\eta')} e^{-ik\mu(\eta-\eta')} \, g(\eta') ( 1 - \mu^2) G_k(\eta'),
\end{equation}
 where
\begin{equation}
 \frac{\partial}{\partial\eta} \kappa(\eta,\eta') = g(\eta),
\quad\mbox{and}\quad
 \kappa(\eta,\eta) = 0.
\end{equation}
By using the formula
\begin{equation}
 e^{ikr\mu}
  = \sum_{m=0}^\infty (2m+1) i^m j_m(kr) P_m(\mu),
\end{equation}
 where $j_\ell(x)$ is the spherical Bessel function, the solution becomes
\begin{eqnarray}
 \beta_{k,\ell} &=& \int_0^\eta d\eta' \, e^{-\kappa(\eta,\eta')} \, g(\eta') \, G_k(\eta')
  i^\ell
  \Bigg\{
   \frac{\ell(\ell-1)}{(2\ell+1)(2\ell-1)} j_{\ell-2}(k(\eta'-\eta))
\nonumber\\
&&
   + \frac{2(\ell^2+\ell-1)}{(2\ell+3)(2\ell-1)} j_{\ell}(k(\eta'-\eta))
   + \frac{(\ell+2)(\ell+1)}{(2\ell+3)(2\ell+1)} j_{\ell+2}(k(\eta'-\eta))
  \Bigg\}.
\label{beta-E-mode}
\end{eqnarray}
This form indicates that $\beta_{k,\ell}$ can be obtained, once the source function has been given.

In general two kinds of polarizations, E-mode and B-mode, are originated from the fact that
 the polarization tensor (here, all its components are small perturbations of the CMB)
 is decomposed in corresponding two scalar components as
\begin{equation}
 P_{ab}(\hat{n}) = 
  \frac{1}{2}
  \left(
   \begin{array}{cc}
    Q & - U \sin\theta \\
    -U \sin\theta & -Q \sin^2\theta
   \end{array}
  \right)
  =
  \sum_{\ell=2}^\infty \sum_{m=-\ell}^\ell
  \left[ a^E_{\ell m} Y^E_{(\ell m)ab}(\hat{n}) + a^B_{\ell m} Y^B_{(\ell m)ab}(\hat{n}) \right],
\end{equation}
 where $Y^E_{(\ell m)ab}(\hat{n})$ and $Y^B_{(\ell m)ab}(\hat{n})$
 are two independent spherical harmonic functions on the sphere
 (see \cite{Cabella:2004mk} for a review).
The coefficients $a^E_{\ell m}$ and $a^B_{\ell m}$ are given as
\begin{eqnarray}
 a^E_{\ell m} &=& \sqrt{\frac{2(\ell-2)!}{(\ell+2)!}}
  \int d\hat{n} \, \nabla_a \nabla_b P^{ab}(\hat{n}) \, ( Y^m_\ell(\hat{n}) )^*,
\nonumber\\
 a^B_{\ell m} &=& \sqrt{\frac{2(\ell-2)!}{(\ell+2)!}}
  \int d\hat{n} \, \nabla_a \nabla_c P^{ab}(\hat{n}) \epsilon^c{}_b\, ( Y^m_\ell(\hat{n}) )^*,
\end{eqnarray}
 where $d\hat{n} = d\theta \sin\theta d\varphi$,
 $\nabla_a$ is the covariant derivative on the sphere,
 $Y^m_\ell(\hat{n})$ is the spherical harmonic function.
If we consider the component $Q$ as a perturbation
 that follows eq.(\ref{perturbation-assumption}) with an additional factor $2$,
 we have $Q_k = T_0 \beta_k$ in the unit of temperature from $Q=I_\theta-I_\varphi$, where
\begin{equation}
 Q(x) = \int \frac{d^3k}{(2\pi)^3}
  \left(
   \hat{\alpha}_{\bf k} Q_k e^{-i{\bf k} \cdot {\bf x}}
   + \hat{\alpha}_{\bf k}^\dag (Q_k)^* e^{i{\bf k} \cdot {\bf x}}
  \right).
\end{equation}
In the present background of scalar perturbation, ${\cal R}^0_k e^{-ikz}$,
 we can describe the component $a^E_{k, \ell m}$ by $\beta_{k,\ell}$ as
\begin{eqnarray}
 a^E_{k, \ell m}
  = &-& \delta_{m0} T_0 \sqrt{\pi(2\ell+1)} \sqrt{\frac{2\ell(\ell-1)}{(\ell+1)(\ell+2)}} \beta_{k,\ell}
\nonumber\\
  &+& \delta_{m0} T_0 \sqrt{\pi(2\ell+1)} \sqrt{\frac{2\ell(\ell-1)}{(\ell+1)(\ell+2)}}
      \sum_{n=1}^{[\ell/2]} \frac{2(\ell-2n)+1}{\ell(\ell-1)/2} \beta_{k,\ell-2n},
\label{alm-E}
\end{eqnarray}
 where we have made a resummation
 by which the infinite summation becomes that of the finite range \cite{Kamionkowski:1996ks}.
The angular power spectrum for the E-mode polarization is given by
\begin{equation}
 C^{EE}_\ell
  = \int \frac{d^3k}{(2\pi)^3} \, \frac{1}{2\ell+1} \sum_{m=-\ell}^\ell \vert a^E_{k,\ell m} \vert^2,
\label{Cl-E}
\end{equation}
 where we assume the isotropy of the universe.

Here, we consider long wavelength limit
 in which we neglect the terms with $-ik$ in equations for $\alpha_{k,\ell}$ and $\beta_{k,\ell}$.
\begin{eqnarray}
 \dot{\alpha}_{k,0} 
 &\simeq& - \frac{{\cal R}^0_k}{45} \eta k^2,
\\
 \dot{\alpha}_{k,1}
 &\simeq& - g \alpha_{k,1},
\\
 \dot{\alpha}_{k,2}
 &\simeq& - \frac{2}{5}\frac{{\cal R}^0_k}{45} \eta k^2
     - g \frac{1}{10} \left( 9 \alpha_{k,2} + \beta_{k,0} - \beta_{k,2} \right),
\\
 \dot{\alpha}_{k,\ell}
 &\simeq& - g \alpha_{k,\ell},
 \quad \ell \geq 3,
\\
 \dot{\beta}_{k,0} 
 &\simeq& - g \frac{1}{2} \left( \beta_{k,0} + \alpha_{k,2} + \beta_{k,2} \right),
\\
 \dot{\beta}_{k,1}
 &\simeq& - g \beta_{k,1},
\\
 \dot{\beta}_{k,2}
 &\simeq& - g \frac{1}{10} \left( 9 \beta_{k,2} + \beta_{k,0} - \alpha_{k,2} \right)
\\
 \dot{\beta}_{k,\ell}
 &\simeq& - g \beta_{k,\ell},
 \quad \ell \geq 3.
\end{eqnarray}
Moreover, neglect $\alpha_{k,\ell}$ and $\beta_{k,\ell}$ of $\ell=1$ and $\ell \geq 3$
 considering that they are all exponentially smaller than the others (so called tight coupling limit),
 then we can obtain a differential equation of the source function
\begin{equation}
 \dot{G}_k = - \frac{3}{10} \, g \, G_k + \frac{{\cal R}^0_k}{150} \, k^2 \eta.
\end{equation}
The solution of this equation can be formally written as
\begin{equation}
 G_k \simeq \frac{1}{10} \int_0^\eta d\eta' e^{-\frac{3}{10}\kappa(\eta,\eta')}
            \frac{{\cal R}^0_k}{15} k^2 \eta'.
\end{equation}
Since we set that the source function vanishes at the time when the reionization starts, $\eta_{\rm ion}$,
 and the exponential factor in the integrant is almost always unity
 in good approximation \cite{Kitazawa:2019fzc},
 we obtain an approximate source function
\begin{equation}
 G_k \simeq \frac{{\cal R}^0_k}{300} k^2 (\eta^2 - \eta_{\rm ion}^2 ).
\end{equation}
Then, from eq.(\ref{beta-E-mode}) in the same approximation we obtain
\begin{eqnarray}
 \beta_{k,\ell} &\simeq& \int_{\eta_{\rm ion}}^\eta d\eta' \, g(\eta') \,
  \frac{{\cal R}^0_k}{300} k^2 (\eta'^2 - \eta_{\rm ion}^2 ) \,
  i^\ell
  \Bigg\{
   \frac{\ell(\ell-1)}{(2\ell+1)(2\ell-1)} j_{\ell-2}(k(\eta'-\eta))
\nonumber\\
&&
   + \frac{2(\ell^2+\ell-1)}{(2\ell+3)(2\ell-1)} j_{\ell}(k(\eta'-\eta))
   + \frac{(\ell+2)(\ell+1)}{(2\ell+3)(2\ell+1)} j_{\ell+2}(k(\eta'-\eta))
  \Bigg\}.
\label{beta-E}
\end{eqnarray}
In this way
 we obtain an approximate formula for E-mode angular power spectrum
 from this result and eqs.(\ref{alm-E}) and (\ref{Cl-E}).
We have used the approximation to obtain $\beta_{k,\ell}$ through the approximate source function
 without fully solving the equations for all $\beta_{k,\ell}$.
 
Exactly the same story applies
 to the B-mode angular power spectrum by tensor perturbations
 (see \cite{Kitazawa:2019fzc} for details).
Here, we simply give results.
The B-mode angular power spectrum due to tensor perturbations is given as
\begin{equation}
 C^{\rm BB}_\ell
  = 2 \int \frac{d^3k}{(2\pi)^3} \frac{1}{2\ell+1} \sum_m | a_{\ell m}^{\rm B}(k) |^2
  = T_0^2 \frac{4}{\pi}
    \int dk k^2
    \left\vert
     \frac{\ell+2}{2\ell+1} \beta_{k, \ell-1}
     +
     \frac{\ell-1}{2\ell+1} \beta_{k, \ell+1}
    \right\vert^2,
\label{Cl-B-mode}
\end{equation}
 where $\beta_{k,\ell}$ is now approximately given as
\begin{equation}
 \beta_{k,\ell}(\eta_0)
  \simeq \int_{\eta_{\rm ion}}^{\eta_0} d\eta'
         i^\ell j_\ell(k(\eta'-\eta_0)) \, g(\eta') G_k(\eta'),
\label{beta-B}
\end{equation}
\begin{equation}
 G_k(\eta)
  \simeq - \frac{1}{10} \left( D_k(\eta) - D_k({\eta_{\rm ion}}) \right).
\end{equation}
Here, the source function
 is described by the amplitude of tensor perturbations in the metric of eq.(\ref{metric}) as
\begin{equation}
 h_{ij}(x) = \int \frac{d^3k}{(2\pi)^3}
  \sum_{a=+,\times}
  \left(
   \hat{\beta}_{\bf k} e^a_{ij} D_k(\eta) e^{-i{\bf k} \cdot {\bf x}}
   + \hat{\beta}_{\bf k}^\dag (e^a_{ij})^\dag (D_k)^* e^{i{\bf k} \cdot {\bf x}}
  \right),
\end{equation} 
 where $\hat{\beta}_{\bf k}$ is another stochastic variable which satisfies
 $\langle \hat{\beta}_{\bf k} \hat{\beta}_{\bf k'}^\dag \rangle
  = (2\pi)^3 \delta^3({\bf k} - {\bf k'})$ and 
\begin{equation}
 D_k(\eta)
  = \sqrt{\frac{2 \pi^2 A_T}{k^3}} \cdot 3 \sqrt{\frac{\pi}{2}}
    (k\eta)^{-3/2} J_{3/2}(k\eta)
\end{equation}
 which is the solution of the evolution equation of tensor perturbations
 ($J_n(x)$ is the Bessel function)
 corresponding to the initial condition, namely, the power spectrum of the primordial tensor perturbation
\begin{equation}
 P_{\cal T} = A_T \left( \frac{k}{k_{\rm pivot}} \right)^{n_T}
\end{equation}
 with $A_T = r_{TS} A_S$, $r_{TS}=0.03$ and $n_T=0$.
There is no deep reason to choose these values,
 and it is simply a set of test values,
 because primordial tensor perturbation has not yet been observed.

\section{The results and implications}
\label{sec:results}

In this section
 we discuss the angular power spectra of the E-mode and B-mode CMB polarizations
 which are produced by Thomson scattering of CMB photons in the period of reionization.
Our main concern is
 how the power spectra are affected by the difference of the expansion of the universe
 that we have discussed in the first section.
The angular power spectra corresponding to two models
 which are suggested by the standard distance ladder and the inverse distance ladder are calculated,
 respectively, and they are compared with those of the concordance $\Lambda$CDM model.
The essential difference of these two models
 is the way of expanding the universe in the period of reionization.

In the semi-analytic method, which has been introduced in the previous section,
 the difference of the behavior of the function $g(\eta) = \sigma_T n_e(\eta) a(\eta)$ is essential,
 where $\sigma_T$ is the Thomson scattering cross section,
 $n_e(\eta)$ is the time--dependent number density of free electrons,
 and $a(\eta)$ is the scale factor.
It can be understood that 
 this function $g(\eta)$ represents the efficiency of producing polarizations at each time.
Note that $n_e(\eta) \propto a(\eta)^{-3}$ and the optical depth
\begin{equation}
 \tau \simeq \int_{\eta_{\rm ion}}^{\eta_0} d \eta' g(\eta')
\label{optical-depth}
\end{equation}
 has been measured as $\tau = 0.054 \pm 0.007$ by the PLANCK collaboration \cite{Aghanim:2018eyx}.
The behavior of $n_e(\eta) a(\eta)^3$
 shows the genuine evolution of free electron density
 without the effect of dilution by the expansion of the universe.
Though the knowledge of the evolution has not yet been established,
 we know that the reionization is finished until $z=6$
 by the observation of Lyman-$\alpha$ forest and Gunn-Peterson though \cite{Gunn:1965hd},
 and its main process starts around $z \simeq 8$.
We set $\eta_{\rm ion}$ to the time corresponding to the redshift $z_{\rm ion}=8$,
 namely $a(\eta_{\rm ion}) = 1/9$,
 applying the relation $1+z=1/a$ with the normalization $a(\eta_0)=1$,
 which is also the condition to set the value of $\eta_0$,
 the age of the universe.
In this work we simply assume
\begin{equation}
 n_e(\eta) a(\eta)^3
  = n_e^{\rm ion} \tanh\left(20 \, \frac{\eta-\eta_{\rm ion}}{\eta_{\rm ion}}\right),
\end{equation}
 where $\eta \geq \eta_{\rm ion}$ and
 $n_e^{\rm ion}$ is a constant which should be determined by eq.(\ref{optical-depth})
 with $\tau=0.054$.
A numerical number $20$ has been chosen
 so that about $90\%$ of reionization has been finished at $z=6$ in our models.

First, we fix the function in case of the $\Lambda$CDM model
 and calculate the E-mode and B-mode power spectra.
Our concordance $\Lambda$CDM model is defined by eq.(\ref{Hubble-CPL})
 with $w_0=-1$, $w_a=0$, $H_0 = 67.4$ [km/s Mpc], $\Omega_m=0.3$ and $\Omega_{\rm DE}=1-\Omega_m=0.7$.
This is the first order differential equation for the scale factor,
 and we solve it with the initial condition $a(\eta=0)=0$.
\begin{figure}[t]
\centering
\includegraphics[width=50mm]{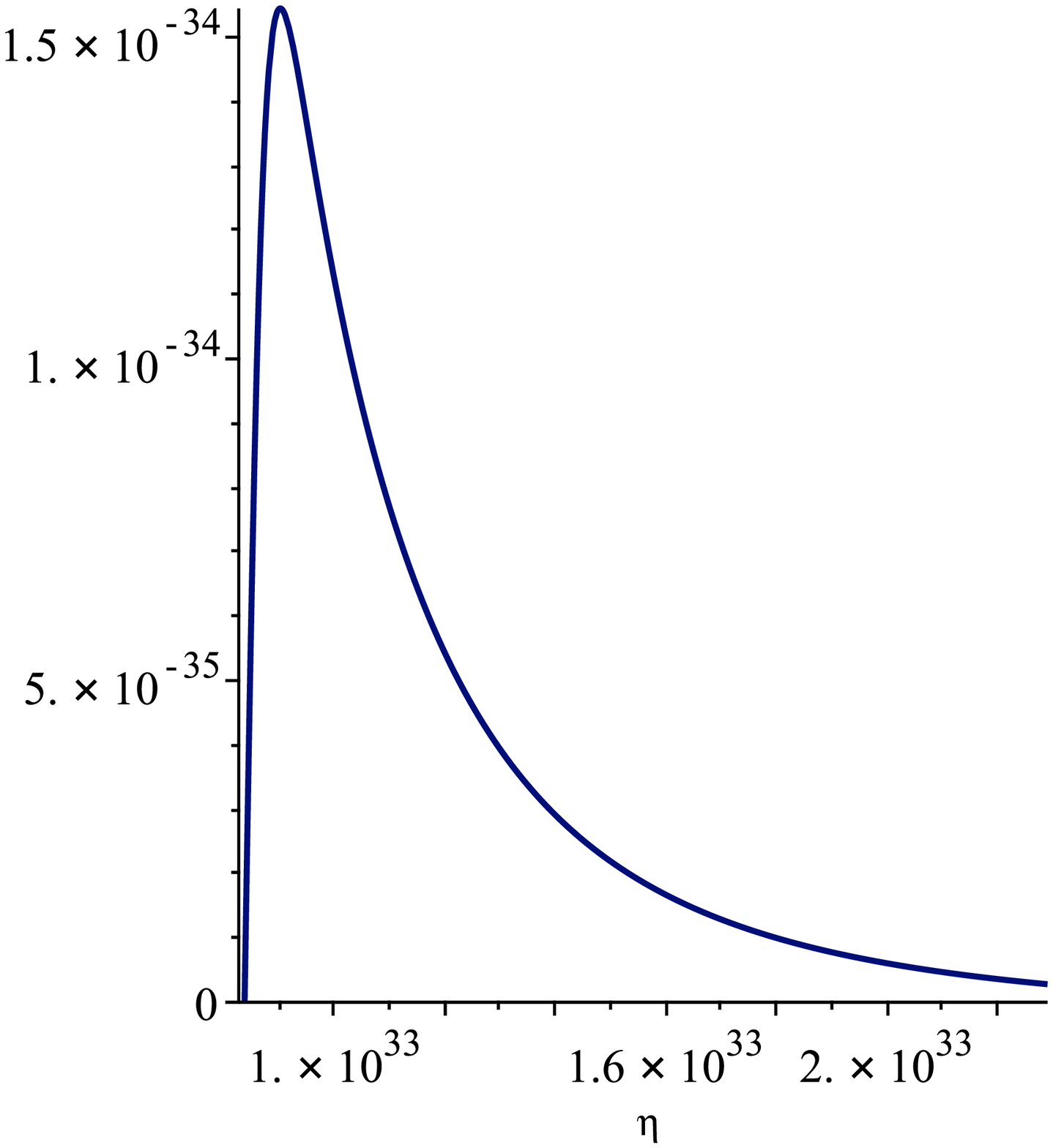}
\hspace{5mm}
\includegraphics[width=50mm]{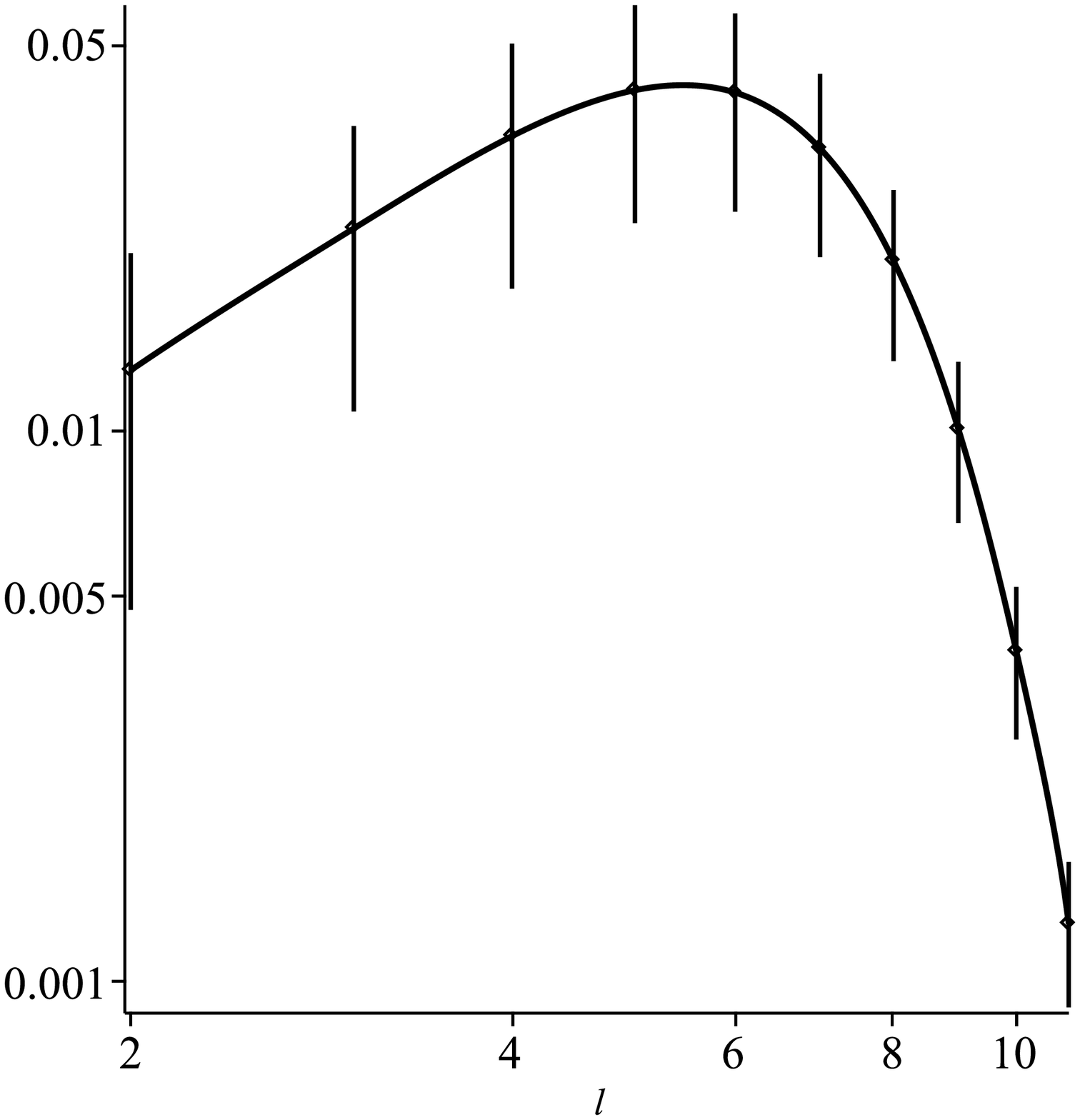}
\hspace{5mm}
\includegraphics[width=50mm]{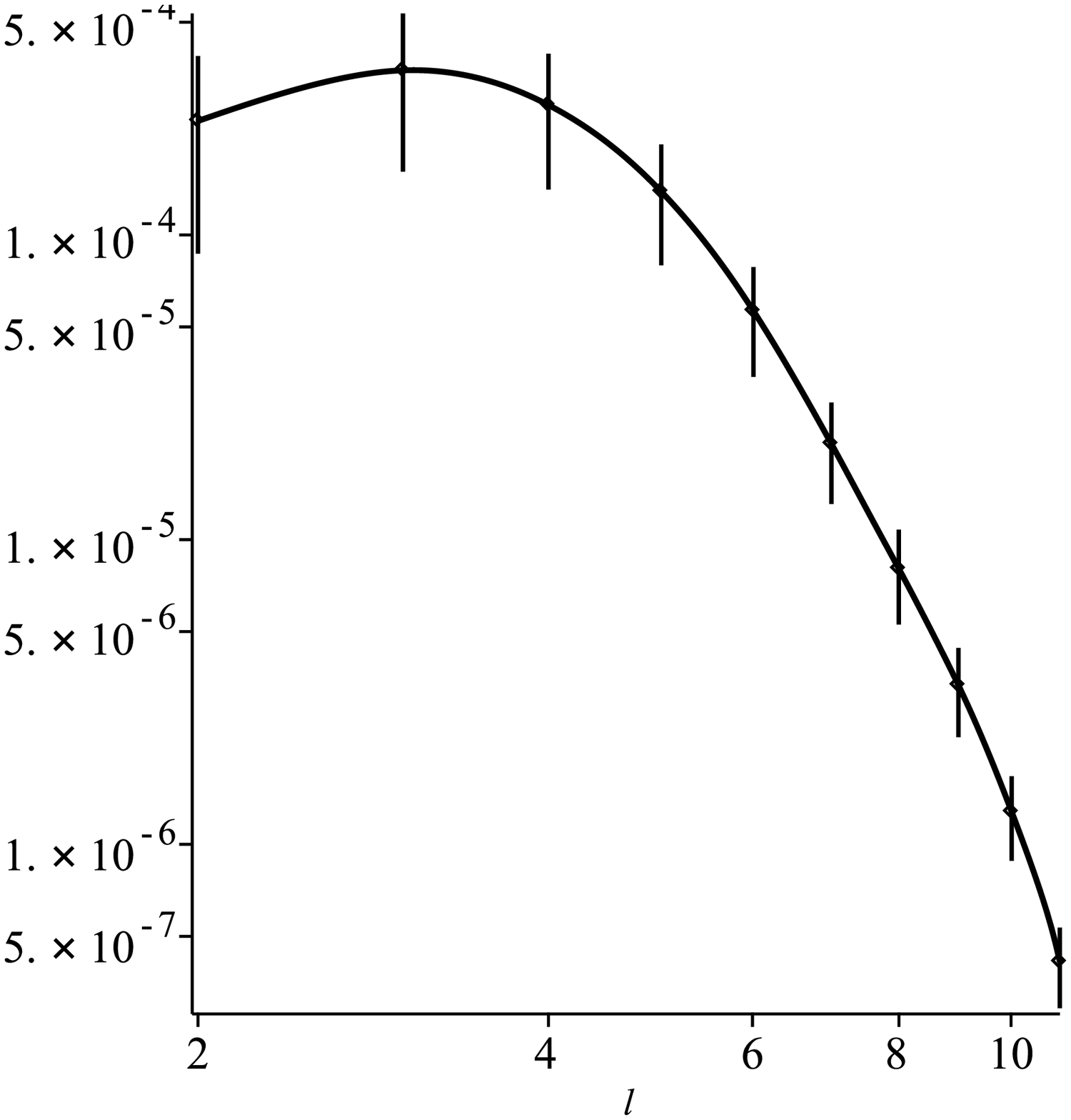}
\caption{
The function $g$ and angular power spectra for the $\Lambda$CDM model.
Left: $g$ [eV] as a function of conformal time $\eta$ [eV${}^{-1}$]
 in the region $\eta_{\rm ion} < \eta < \eta_0$.
Middle: the angular power spectrum of E-mode polarization $D^{EE}_\ell$ [$\mu$K${}^2$]
 with errors by cosmic variance.
Right: the angular power spectrum of B-mode polarization $D^{BB}_\ell$ [$\mu$K${}^2$]
 with errors by cosmic variance.
}
\label{fig:LCDM}
\end{figure}
In fig.\ref{fig:LCDM}
 we present the function $g$ and angular power spectra in the $\Lambda$CDM model.
The plot of the function $g$ indicates that
 the efficiency to produce polarizations quickly increases
 as the increase of the free electron density just after reionization starts,
 and then the efficiency rather quickly decreases
 as the decrease of the free electron density by the expansion of the universe.
Both E-mode and B-mode angular power spectra roughly reflect this behavior:
 larger $\ell$ powers, which have produced at early time, are small,
 and then they become larger for smaller $\ell$,
 but lower $\ell$ powers, which have produced at late time, become small again.
Although our approximation is rather drastic,
 the resultant power spectra are in rather good agreement with more rigorous calculations
 in not only their shapes but also even quantitatively. 
We assume the cosmic variance limited measurement
 and the corresponding errors are included in the plots in fig.\ref{fig:LCDM}. 

\begin{figure}[t]
\centering
\includegraphics[width=40mm]{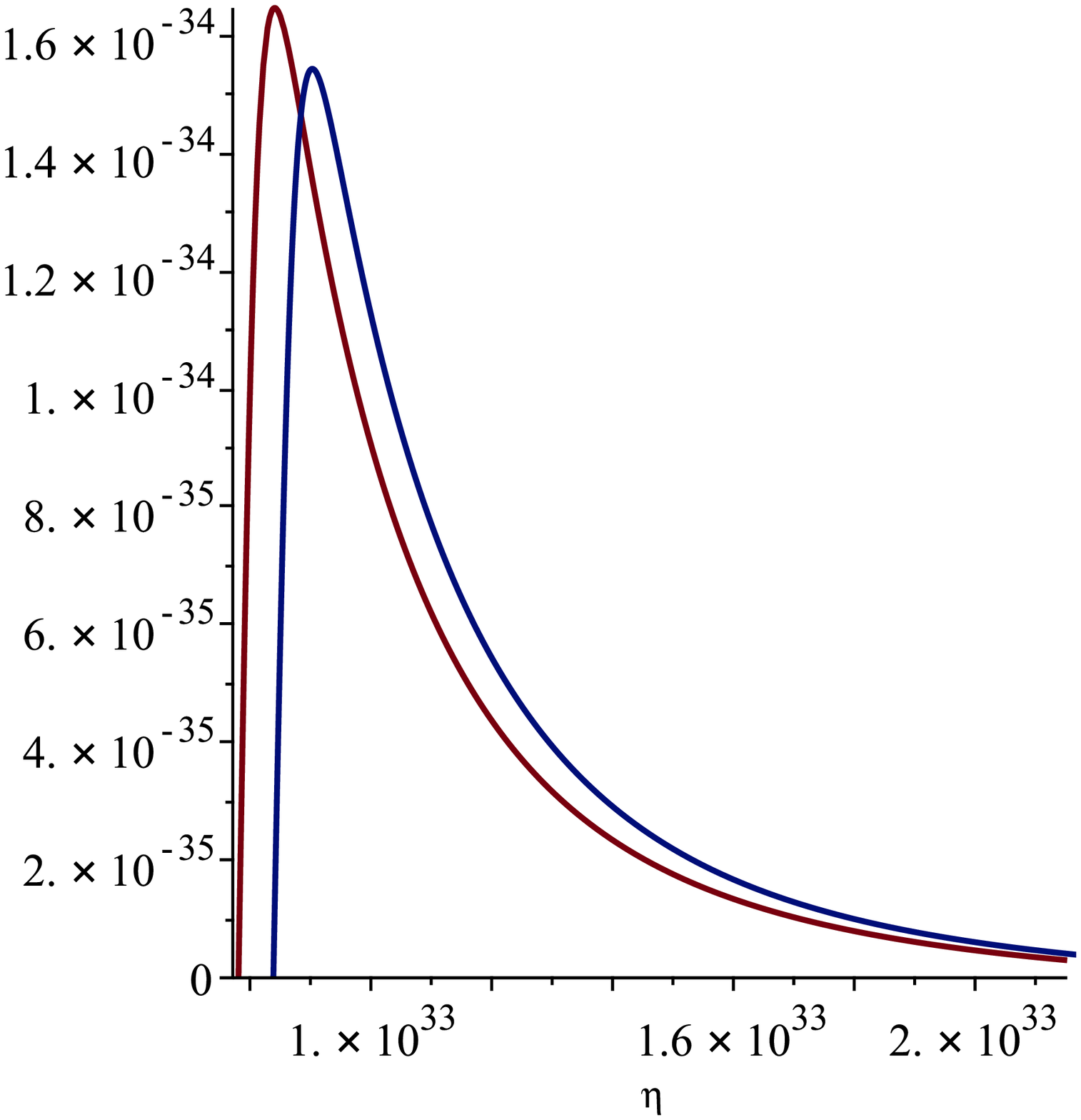}
\hspace{5mm}
\includegraphics[width=40mm]{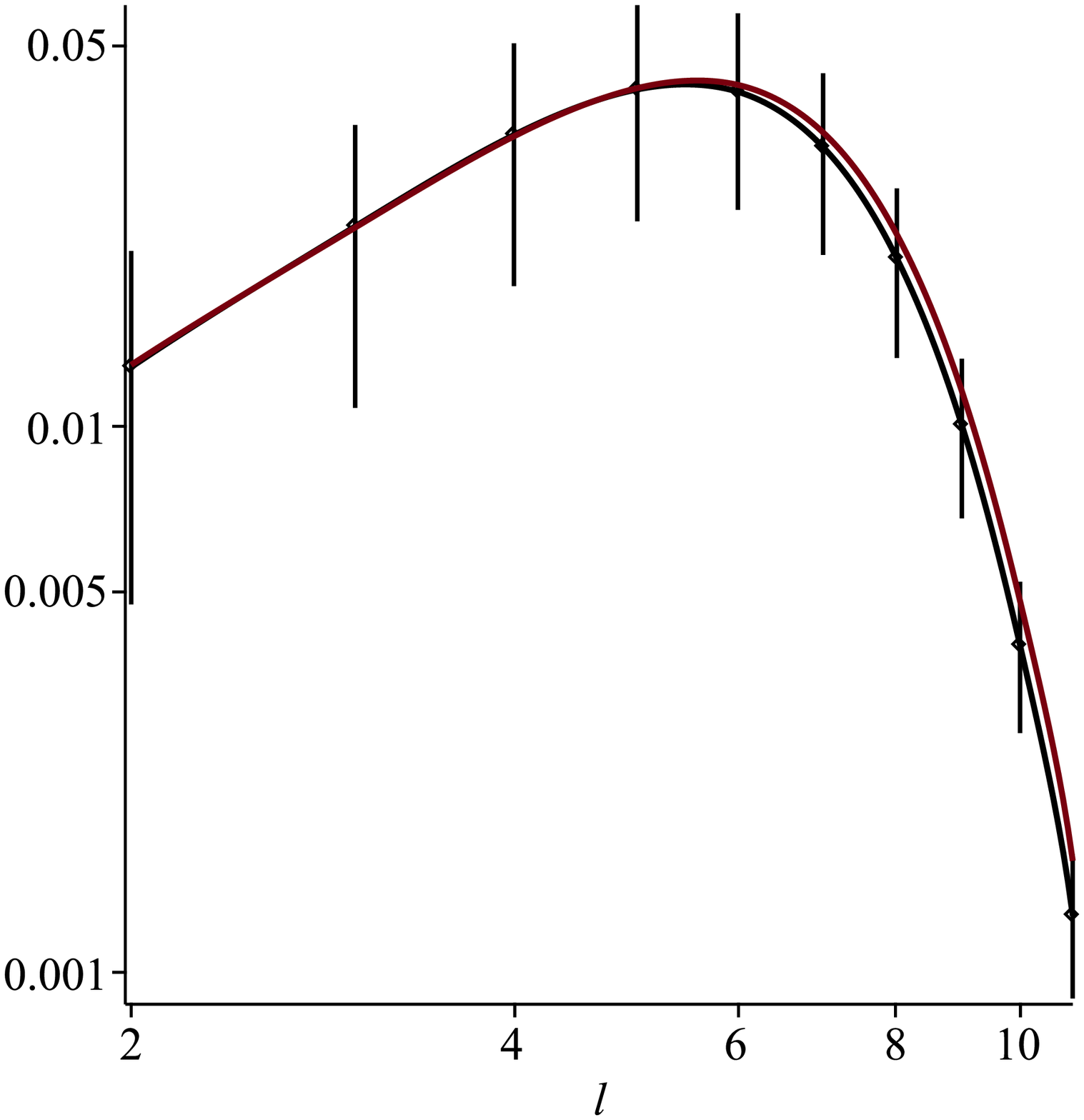}
\hspace{5mm}
\includegraphics[width=40mm]{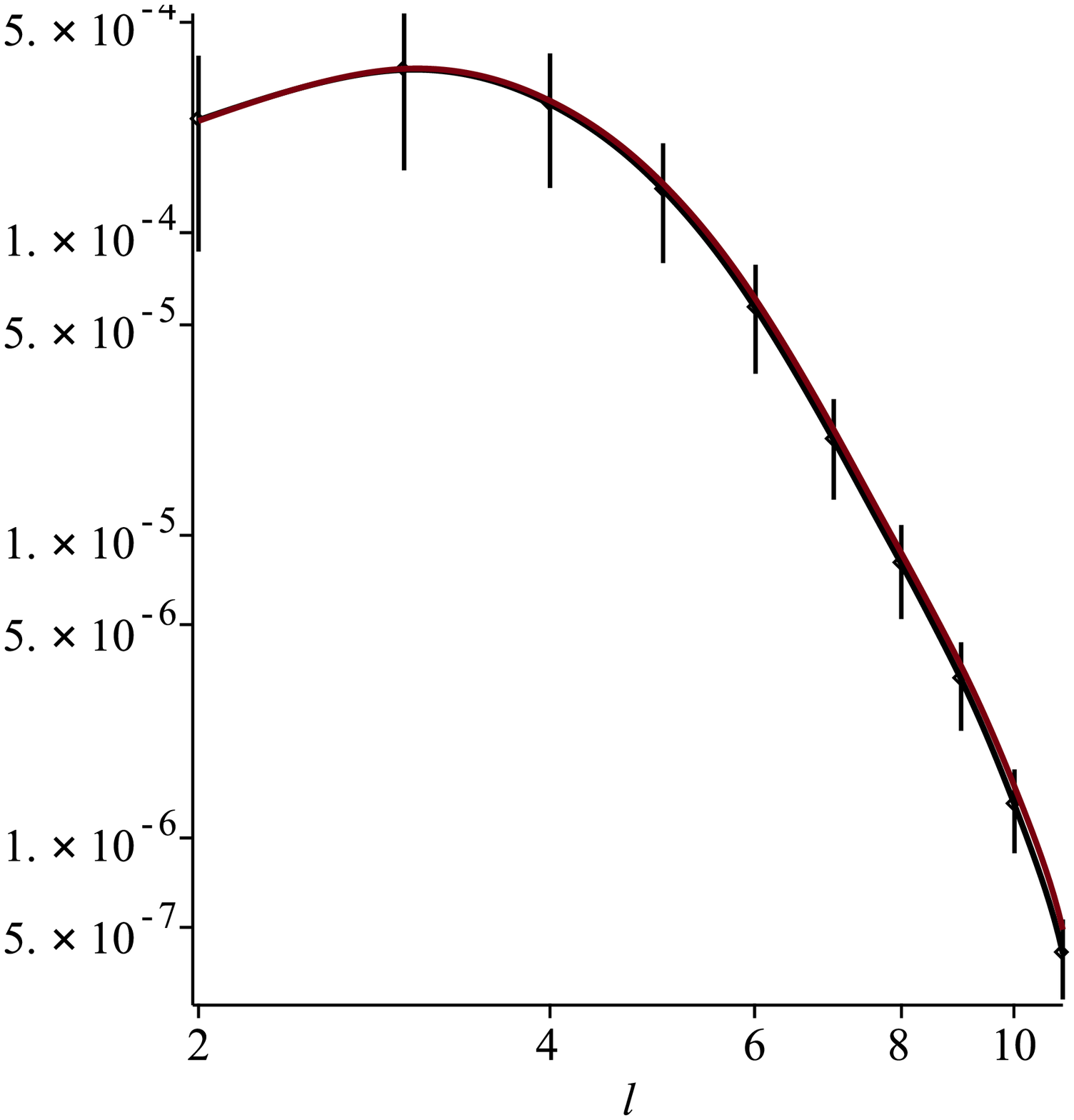}
\caption{
The function $g$ and angular power spectra for the CPL model with the standard distance ladder.
Left: $g$ [eV] as a function of conformal time $\eta$ [eV${}^{-1}$]
 in the region $\eta_{\rm ion} < \eta < \eta_0$.
The line with higher peak is that of this model being compared with that of the $\Lambda$CDM model.
Middle: the angular power spectra of E-mode polarization $D^{EE}_\ell$ [$\mu$K${}^2$]
 with errors by cosmic variance.
The spectrum of this model almost overlaps with that of the $\Lambda$CDM model
 with some excess at high $\ell$.
Right: the angular power spectra of B-mode polarization $D^{BB}_\ell$ [$\mu$K${}^2$]
 with errors by cosmic variance.
The spectrum of this model almost overlaps with that of the $\Lambda$CDM model
 with some excess at high $\ell$.
}
\label{fig:CPL-SH0ES}
\end{figure}
\begin{figure}[t]
\centering
\includegraphics[width=40mm]{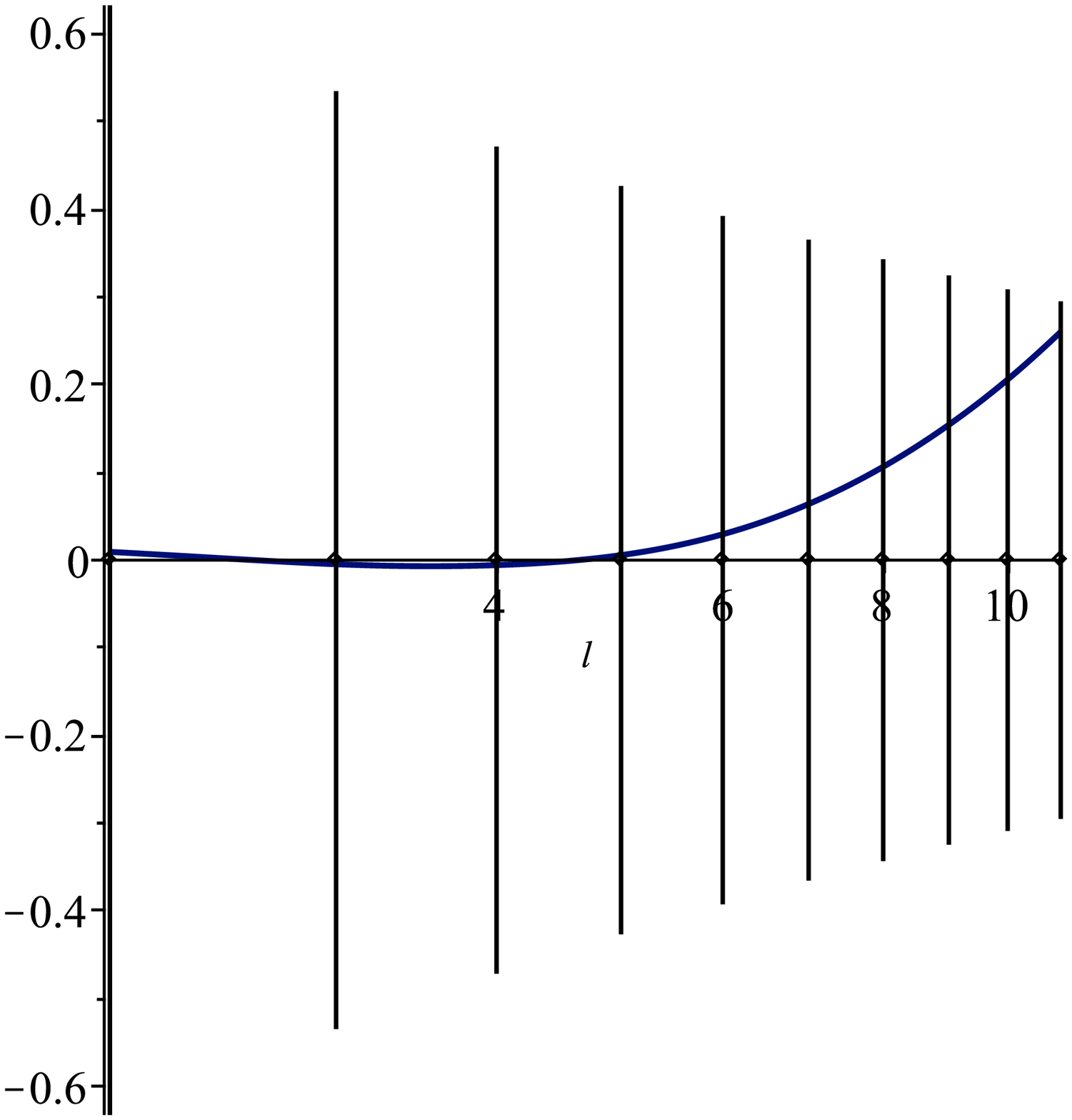}
\hspace{5mm}
\includegraphics[width=40mm]{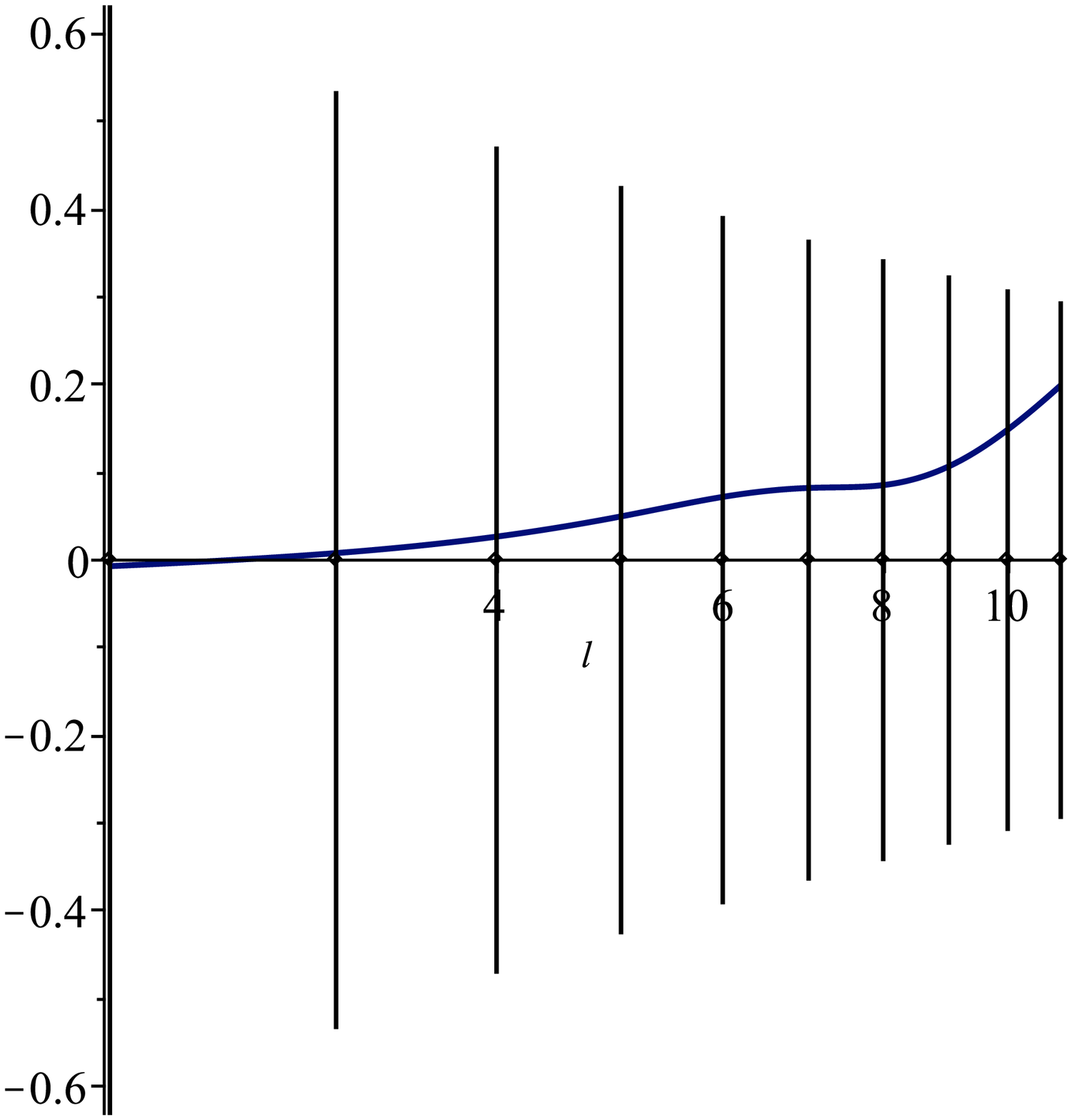}
\caption{
The difference of the angular power spectra in the CPL model with the standard distance ladder
 and that in the $\Lambda$CDM model.
Left: plot of $(D^{EE}_\ell - D^{EE, \Lambda{\rm CDM}}_\ell)/D^{EE, \Lambda{\rm CDM}}_\ell$
 with errors by cosmic variance in vertical lines.
Right: plot of $(D^{BB}_\ell - D^{BB, \Lambda{\rm CDM}}_\ell)/D^{BB, \Lambda{\rm CDM}}_\ell$
 with errors by cosmic variance in vertical lines.
}
\label{fig:CPL-SH0ES-diff}
\end{figure}
In fig.\ref{fig:CPL-SH0ES}
 we present the results in the CPL model with the standard distance ladder.
There is a clear difference in the function $g$ between this model and the $\Lambda$CDM model.
The peak becomes higher and every behaviors become quicker,
 because the time to start reionization is earlier,
 the age of the universe becomes shorter,
 and the period of reionization $\eta_0-\eta_{\rm ion}$ becomes shorter.
Note that the function $g$ is constrained by eq.(\ref{optical-depth}).
As the result
 at larger $\ell$ the values of both $D^{EE}_\ell$ and $D^{BB}_\ell$ are enhanced,
 though the values at smaller $\ell$ are not changed.
This change of the shape of the angular power spectra
 are clearly seen in fig.\ref{fig:CPL-SH0ES-diff}.
Note that
 this magnitude of distortion of the angular power spectra about $20\%$ does not happen
 by simply changing the value of $H_0$ keeping $w=-1$.
For example,
 if we simply change the value of $H_0$ from $67.4$ to $74.0$ [km/s Mpc] with fixed $w=-1$,
 the change of both E-mode and B-mode power spectra is less than $1\%$.

\begin{figure}[t]
\centering
\includegraphics[width=40mm]{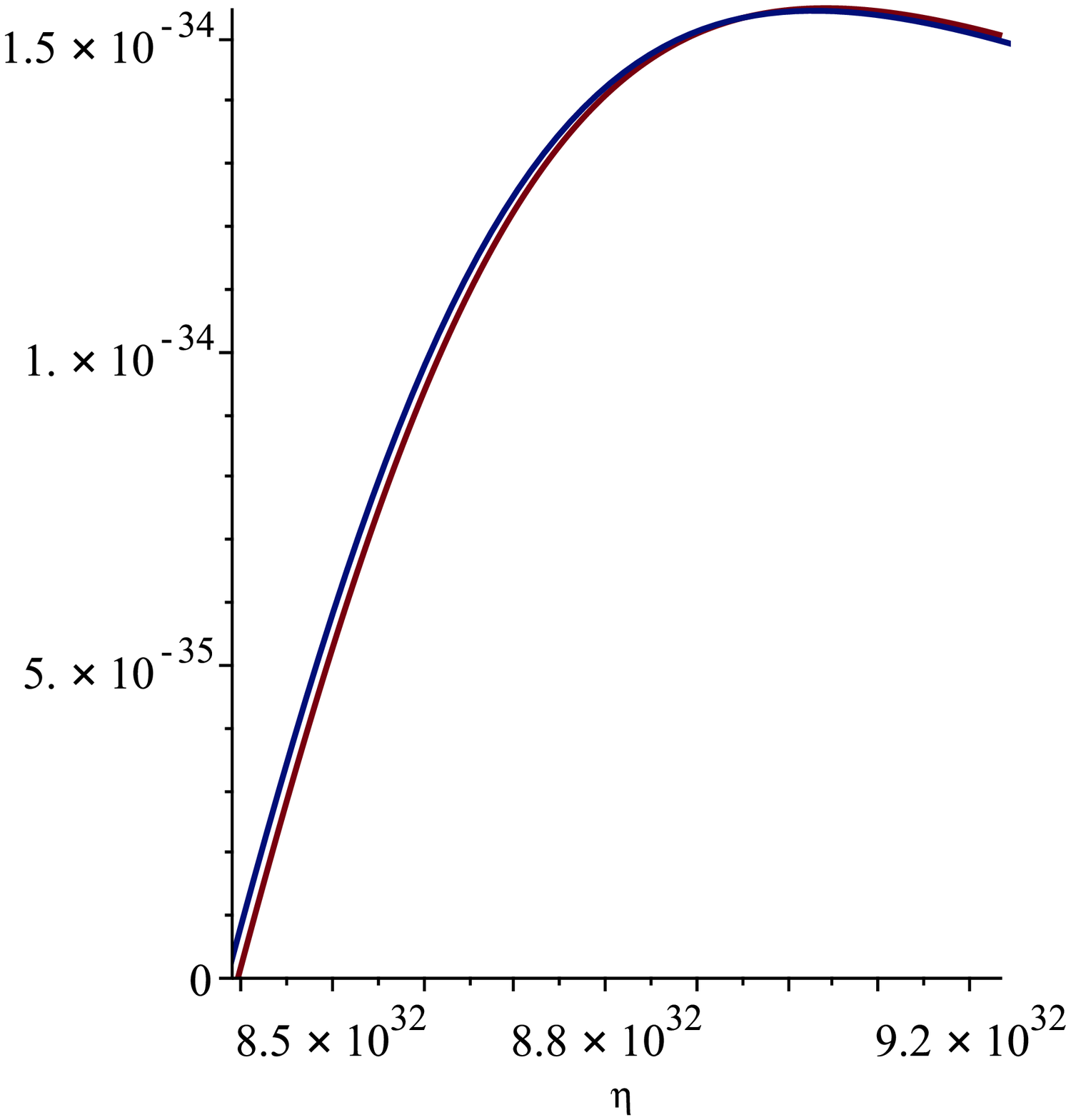}
\hspace{5mm}
\includegraphics[width=40mm]{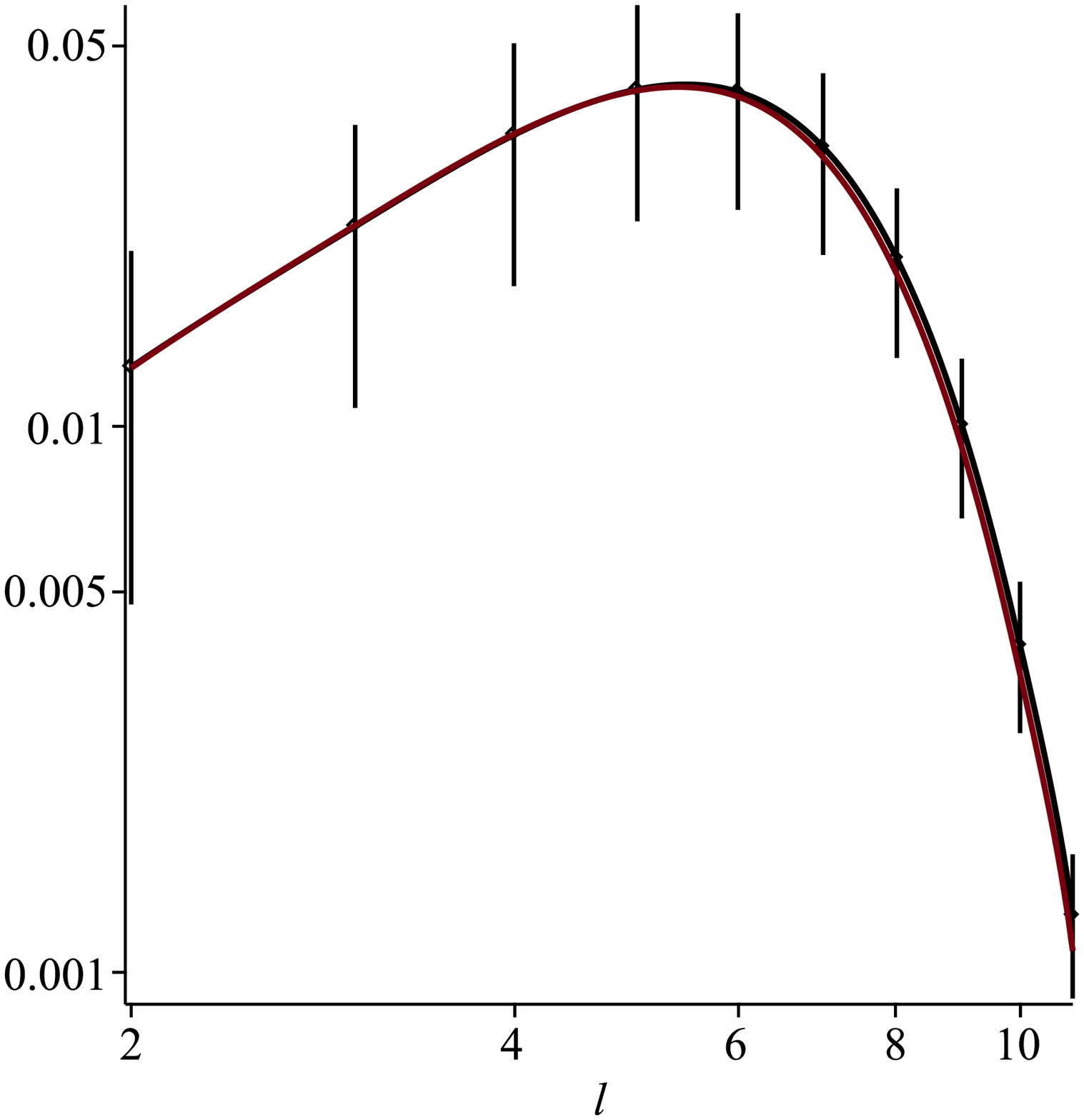}
\hspace{5mm}
\includegraphics[width=40mm]{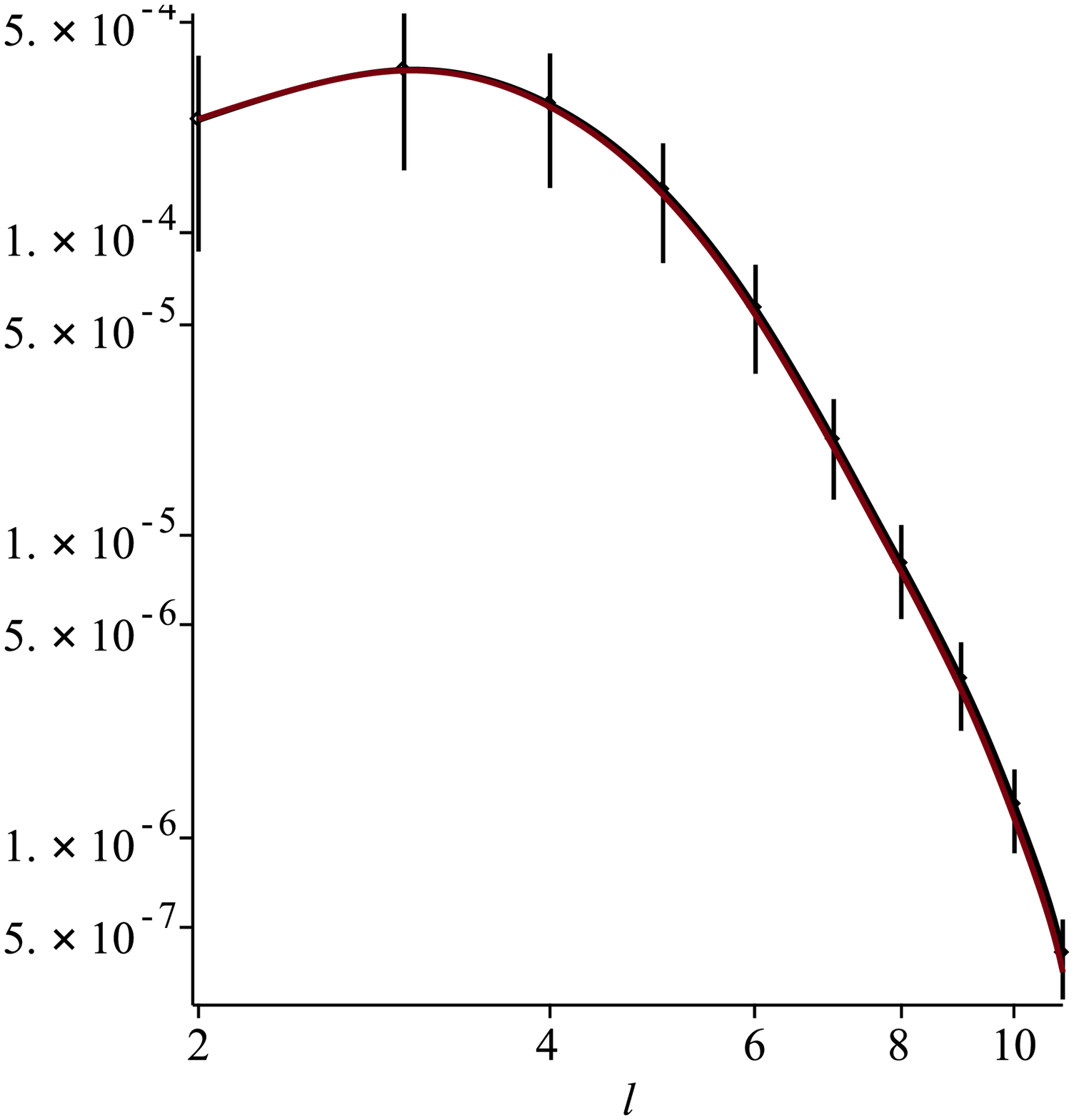}
\caption{
The function $g$ and angular power spectra for the CPL model with the inverse distance ladder.
Left: $g$ [eV] as a function of conformal time $\eta$ [eV${}^{-1}$]
 in the region of $\eta$ close to $\eta_{\rm ion}$.
The lines corresponding to the model with the inverse distance ladder
 and the $\Lambda$CDM model almost overlap with very little shift of the former to the right in $\eta$. 
Middle: the angular power spectra of E-mode polarization $D^{EE}_\ell$ [$\mu$K${}^2$]
 with errors by cosmic variance.
The spectrum of this model almost overlaps with that of the $\Lambda$CDM model
 with some suppression at high $\ell$ (difficult to see with eyes).
Right: the angular power spectra of B-mode polarization $D^{BB}_\ell$ [$\mu$K${}^2$]
 with errors by cosmic variance.
The spectrum of this model almost overlaps with that of the $\Lambda$CDM model
 with some suppression at high $\ell$ (difficult to see with eyes).
}
\label{fig:CPL-BOSS}
\end{figure}
\begin{figure}[t]
\centering
\includegraphics[width=40mm]{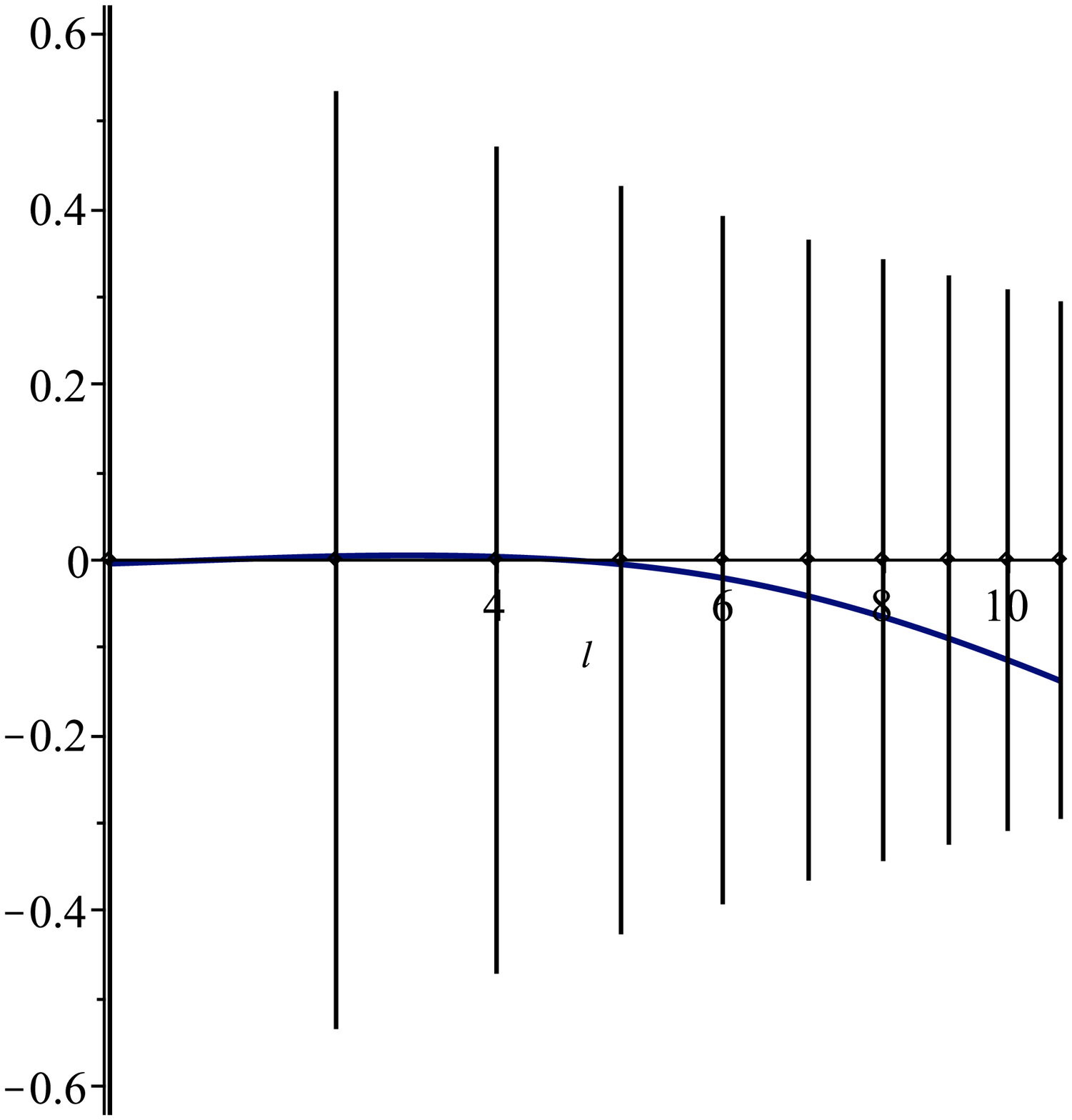}
\hspace{5mm}
\includegraphics[width=40mm]{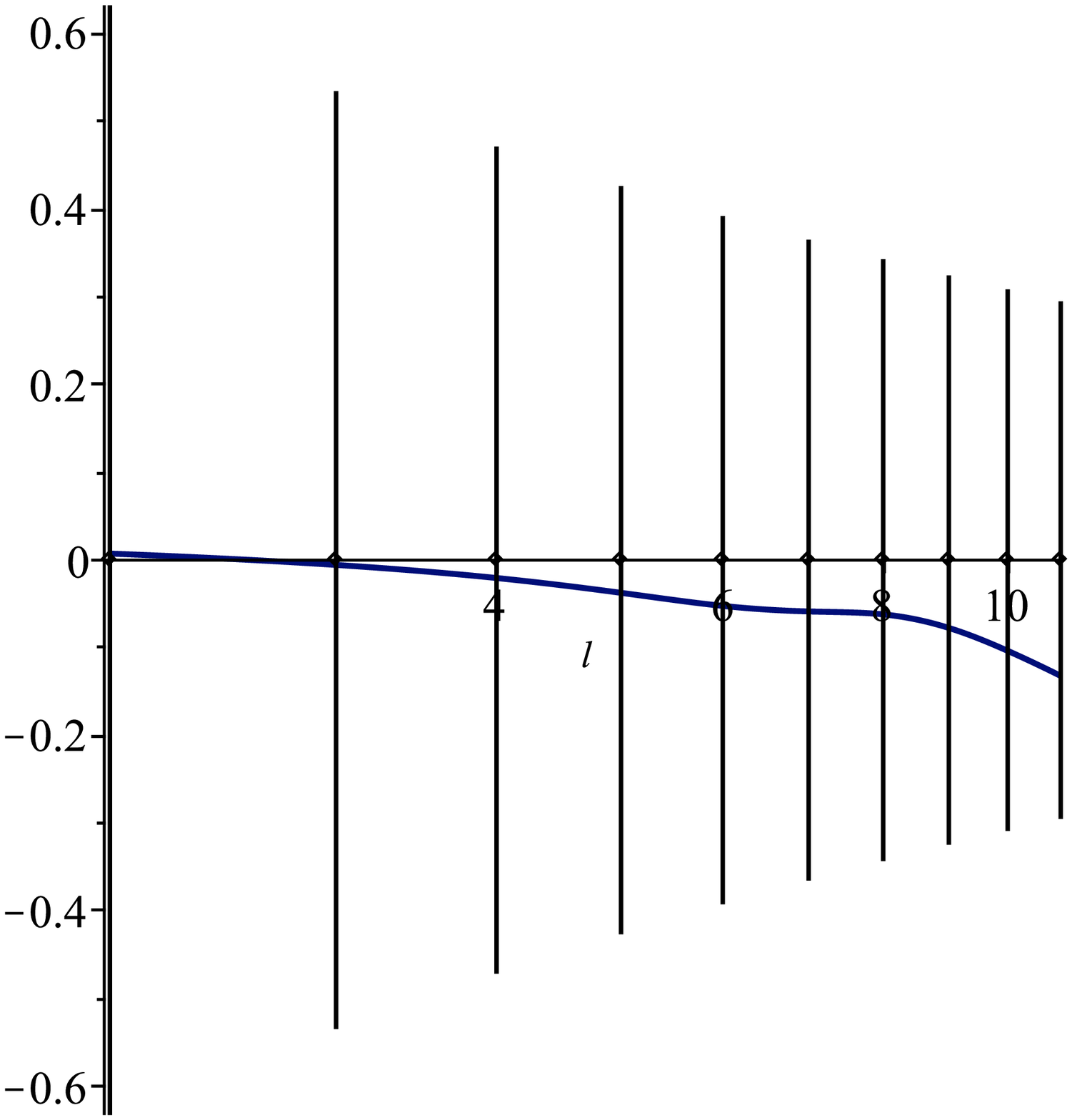}
\caption{
The difference of the angular power spectra in the CPL model with the inverse distance ladder
 and that in the $\Lambda$CDM model.
Left: plot of $(D^{EE}_\ell - D^{EE, \Lambda{\rm CDM}}_\ell)/D^{EE, \Lambda{\rm CDM}}_\ell$
 with errors by cosmic variance in vertical lines.
Right: plot of $(D^{BB}_\ell - D^{BB, \Lambda{\rm CDM}}_\ell)/D^{BB, \Lambda{\rm CDM}}_\ell$
 with errors by cosmic variance in vertical lines.
}
\label{fig:CPL-BOSS-diff}
\end{figure}
In fig.\ref{fig:CPL-BOSS}
 we present the results in the CPL model with the inverse distance ladder.
The difference in the function $g$ is small as expected,
 which comes from that reionization starts slightly later
 and the age of the universe becomes slightly shorter.
Since the difference is little, the change of angular power spectra is very small.
The change can be only seen in the plots of fig.\ref{fig:CPL-BOSS-diff},
 which shows that the powers at larger $\ell$ become smaller.
Note that this behavior is opposite to that in the model with the standard distance ladder.
This little dump comes from slightly later start of reionization.
Since the difference from the $\Lambda$CDM model is small,
 we may interpret this result as that
 the data set of the inverse distance ladder favors the $\Lambda$CDM model,
 or dark energy as a cosmological constant.

\begin{figure}[t]
\centering
\includegraphics[width=40mm]{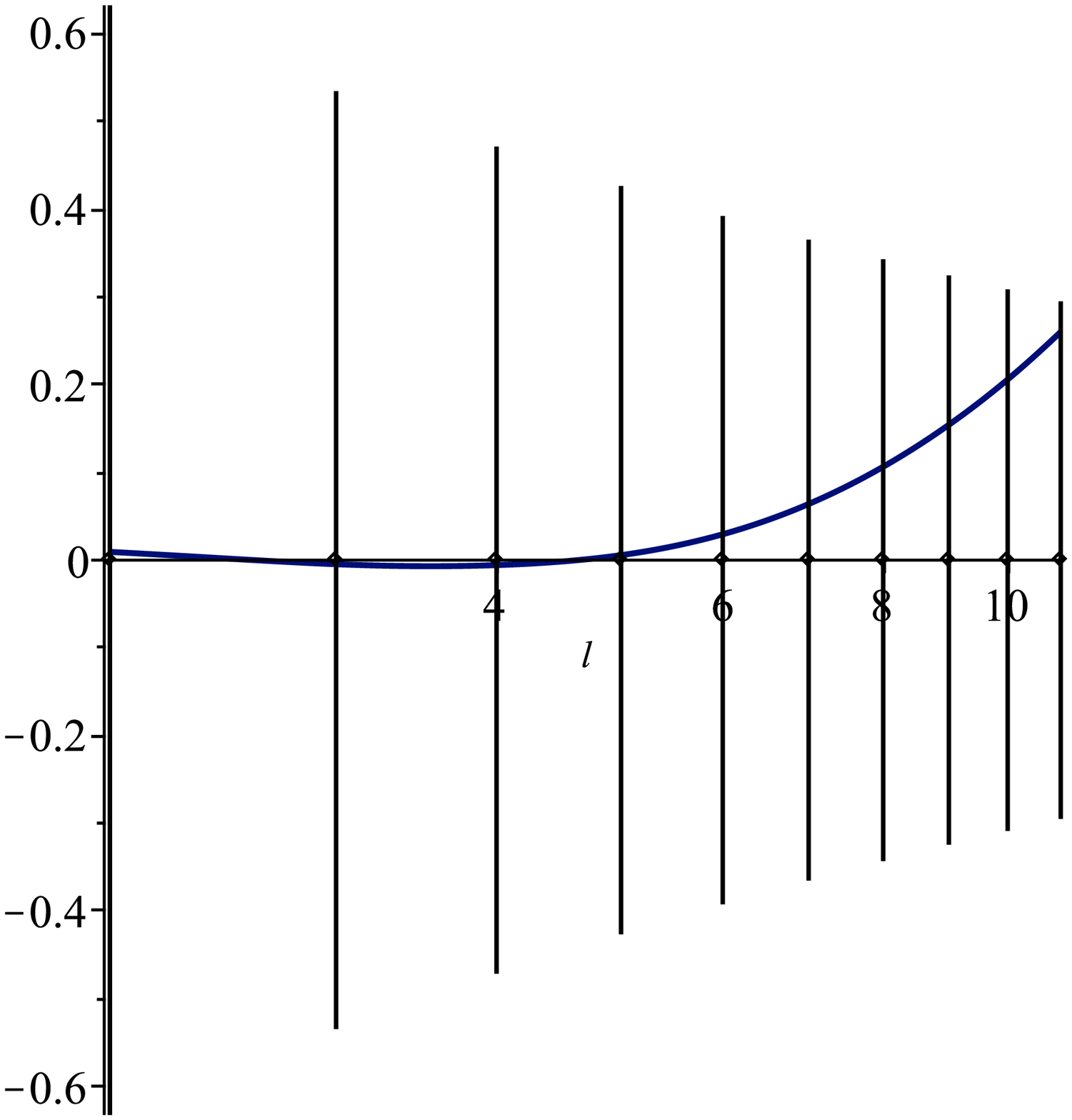}
\includegraphics[width=40mm]{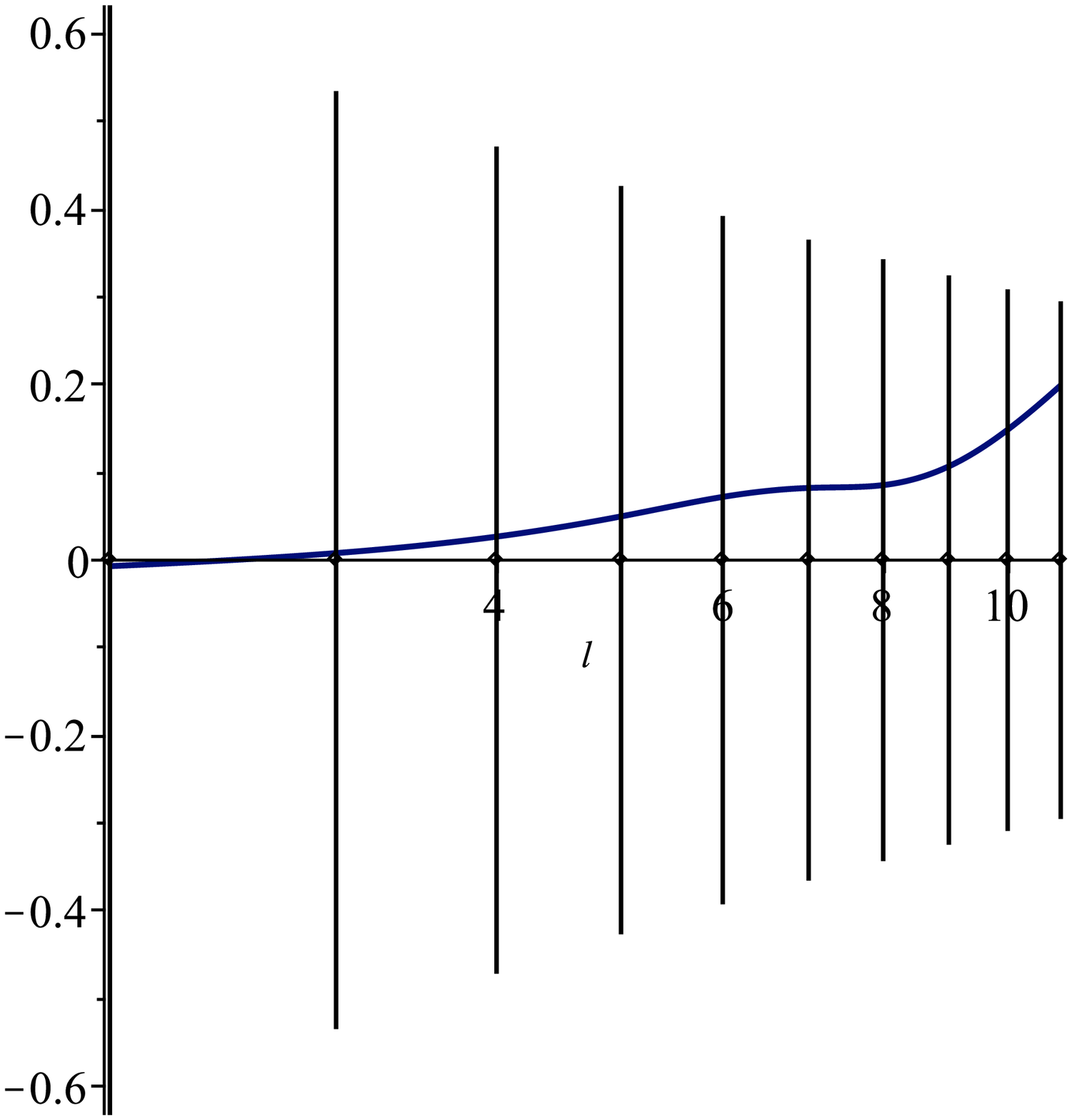}
\includegraphics[width=40mm]{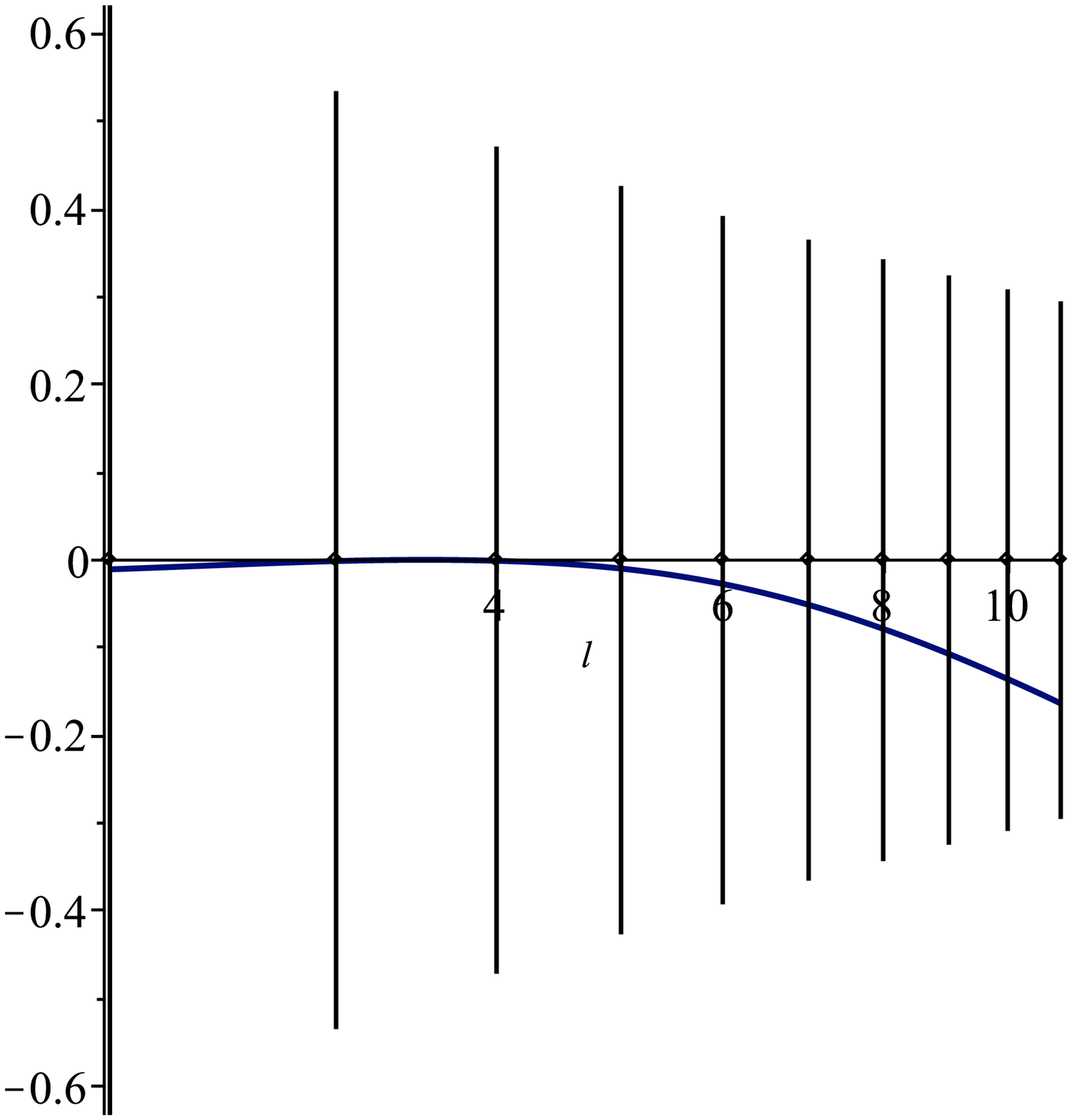}
\includegraphics[width=40mm]{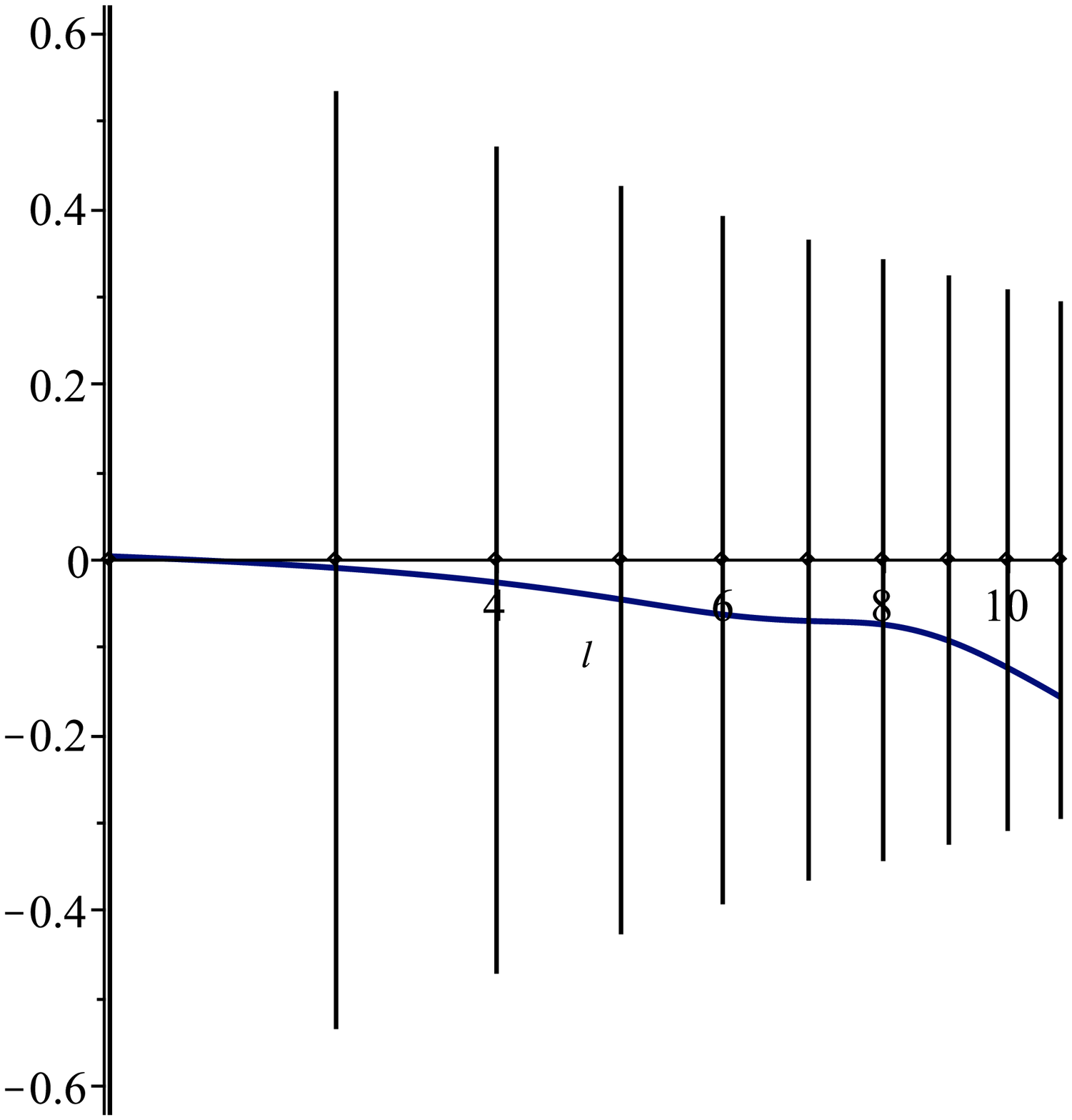}
\caption{
The difference of the angular power spectra in the model of Taylor expansion
 with the standard distance ladder and the inverse distance ladder and that in the $\Lambda$CDM model.
Left: plot of $(D^{EE}_\ell - D^{EE, \Lambda{\rm CDM}}_\ell)/D^{EE, \Lambda{\rm CDM}}_\ell$
 in the model with the standard distance ladder.
Middle-Left: plot of $(D^{BB}_\ell - D^{BB, \Lambda{\rm CDM}}_\ell)/D^{BB, \Lambda{\rm CDM}}_\ell$
 in the model with the standard distance ladder.
Middle-Right: plot of $(D^{EE}_\ell - D^{EE, \Lambda{\rm CDM}}_\ell)/D^{EE, \Lambda{\rm CDM}}_\ell$
 in the model with the inverse distance ladder.
Right: plot of $(D^{BB}_\ell - D^{BB, \Lambda{\rm CDM}}_\ell)/D^{BB, \Lambda{\rm CDM}}_\ell$
 in the model with the inverse distance ladder.
In all plots vertical lines indicate errors by cosmic variance.
}
\label{fig:Taylor-diff}
\end{figure}
Fig.\ref{fig:Taylor-diff}
 shows that we have completely the same result in the models of Taylor expansion,
 which leads a model independent observation:
 in the production of the polarizations of the CMB in the period of reionization
 with non--trivially time--dependent equation of state of dark energy,
 the evolution of the universe suggested by the standard distance ladder
 indicates higher polarization powers at larger $\ell$,
 and the evolution of the universe suggested by the inverse distance ladder
 indicates lower polarization powers at larger $\ell$.
This means that
 the future precise measurements of E-mode and possibly B-mode angular power spectra
 at low-$\ell$ by LiteBIRD, for example,
 can give a useful information to the Hubble tension.
Suppose that we find higher polarization powers in future experiments.
It can be understood that
 the non--trivially time--dependent equation of state of dark energy with the standard distance ladder,
 namely larger value of the Hubble constant with new physics at low-$z$, is favored.

\begin{figure}[t]
\centering
\includegraphics[width=40mm]{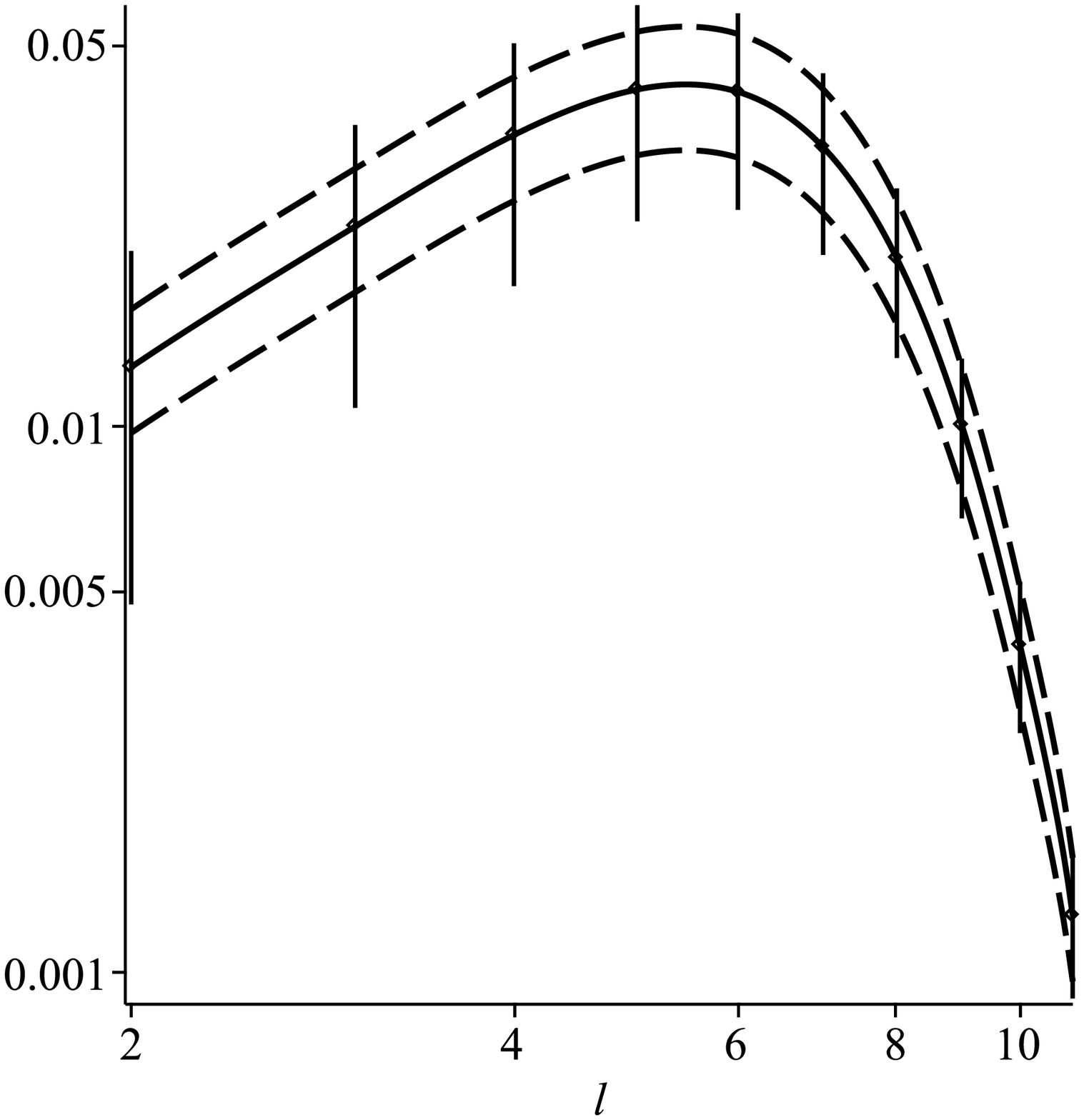}
\includegraphics[width=40mm]{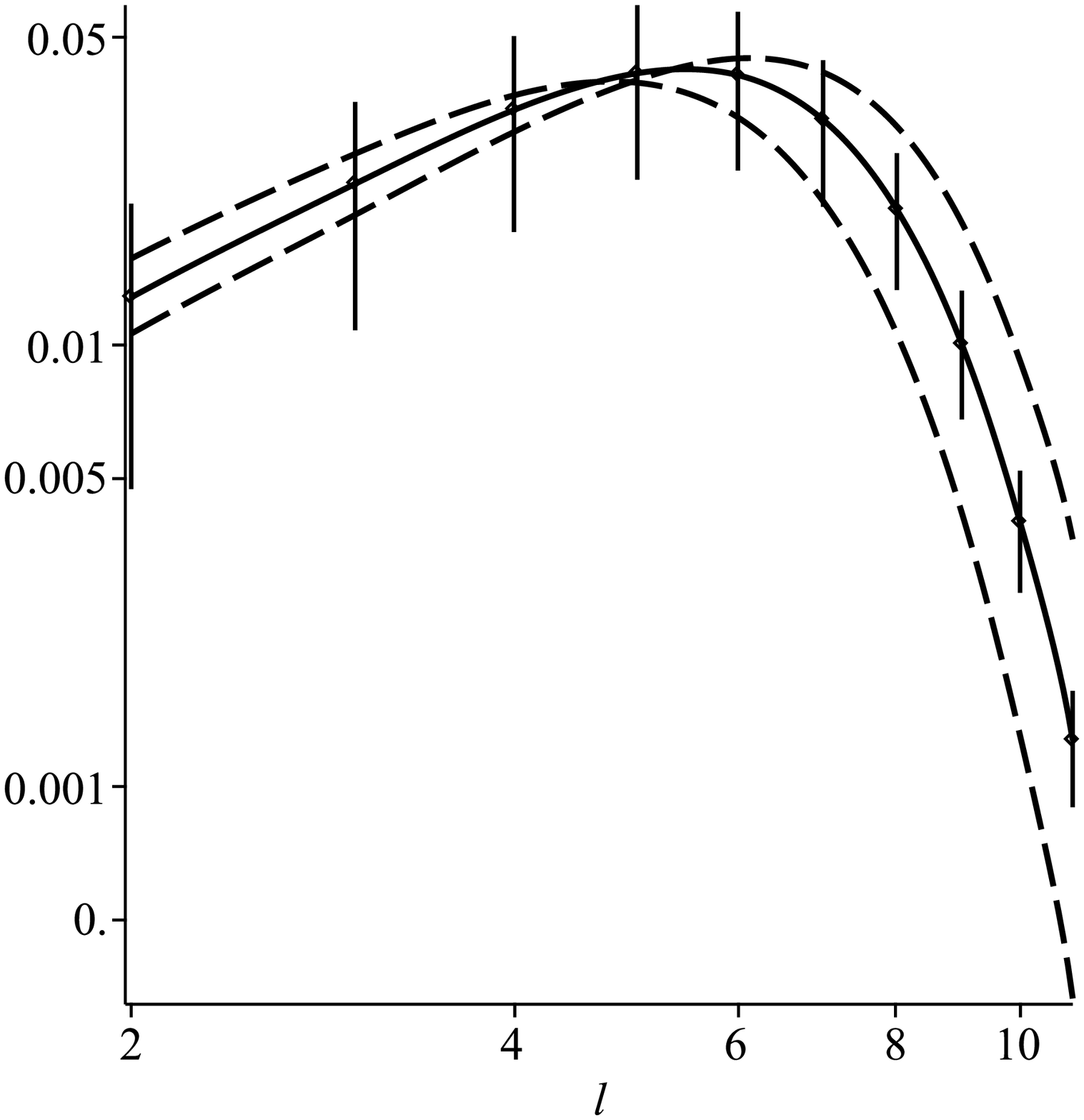}
\includegraphics[width=40mm]{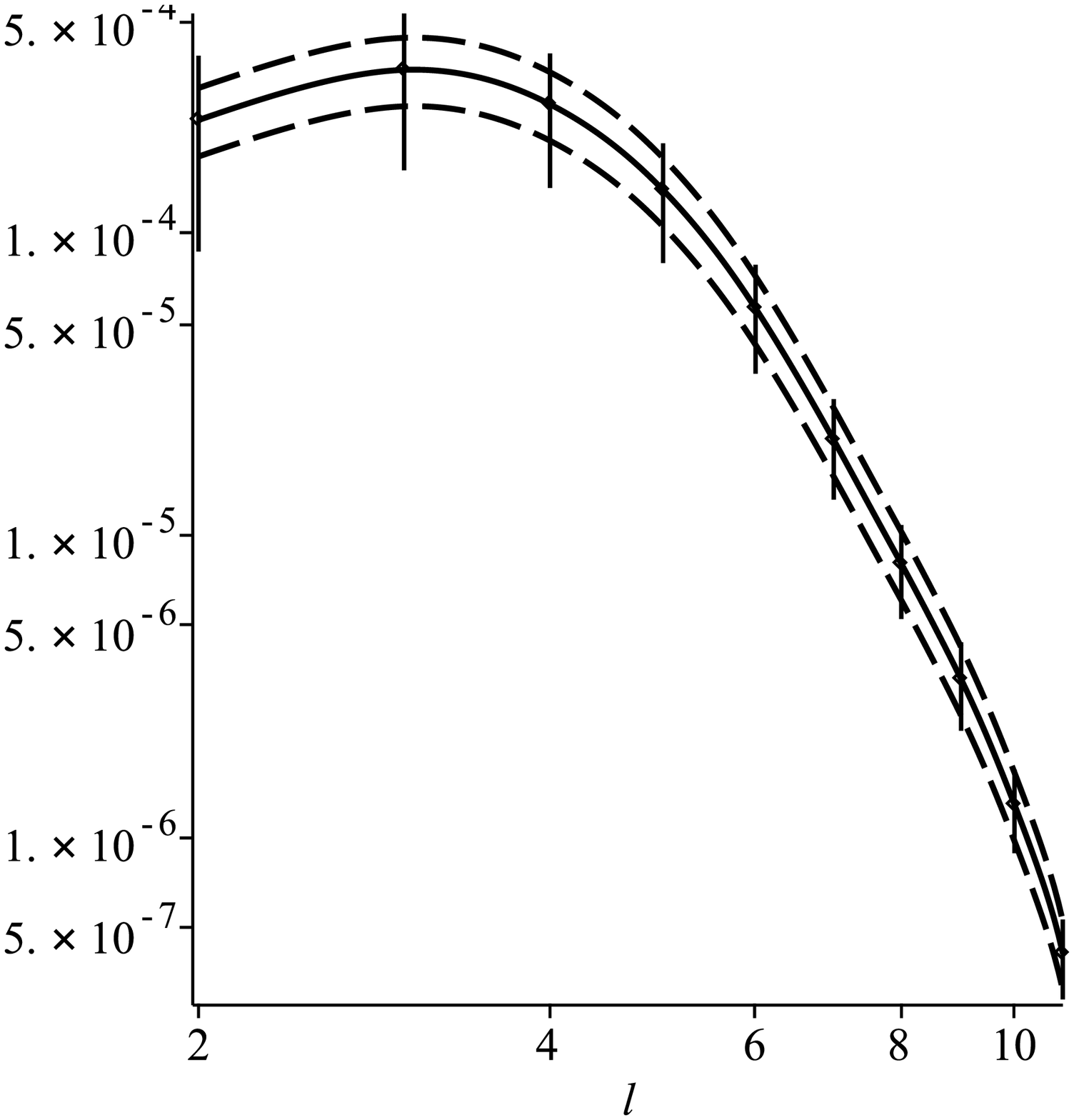}
\includegraphics[width=40mm]{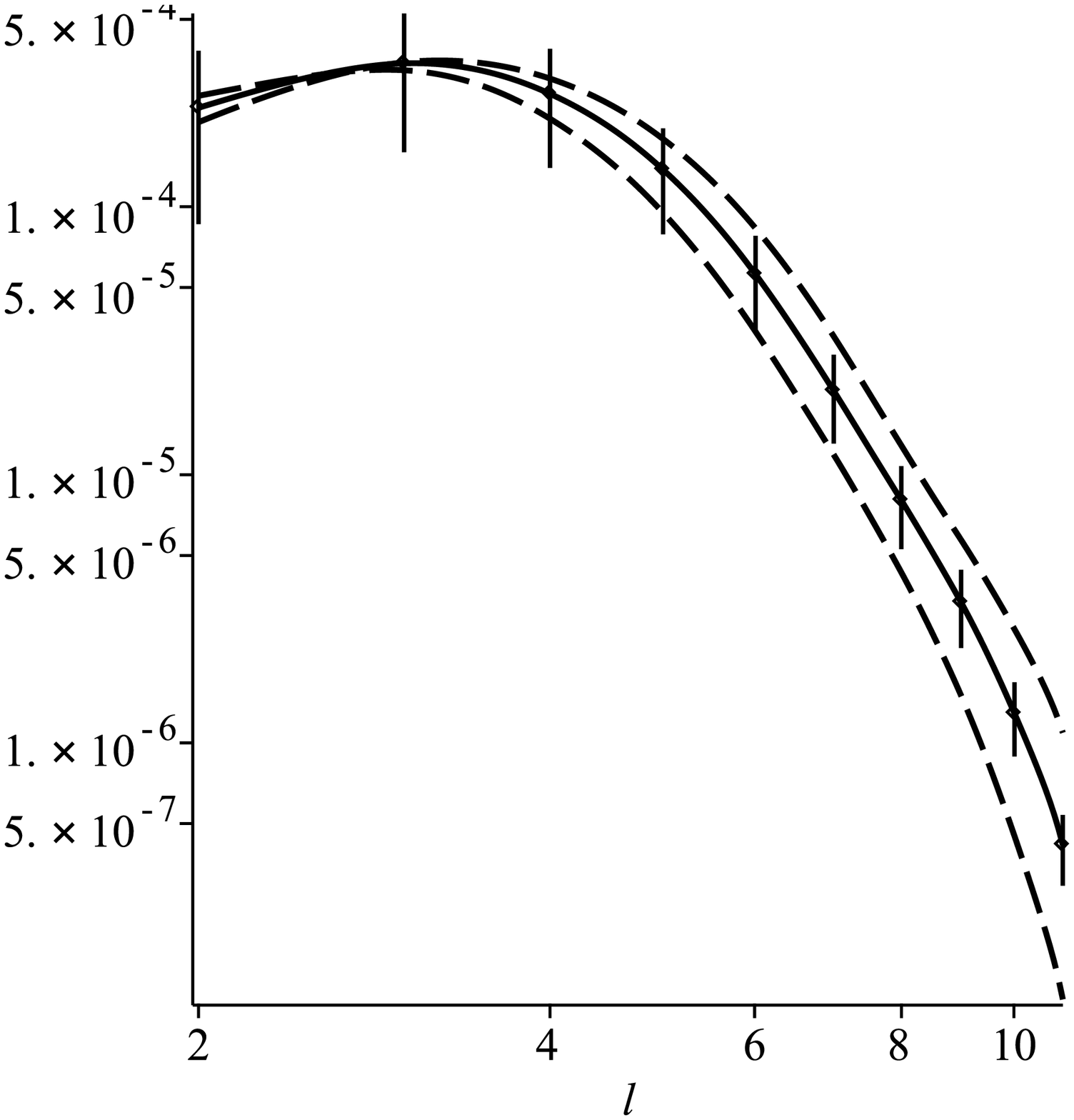}
\caption{
The ambiguities of angular power spectra
 under changing the value of optical depth $\tau$
 and the value of redshift $z_{\rm ion}$ when reionization starts.
Left: $D^{EE}_\ell$ with different values of $\tau = 0.054 \pm 0.007$.
 Three lines are corresponding to those of the upper bound value, center value and lower bound value,
 from up to down, respectively.
Middle-Left: $D^{EE}_\ell$ with different value of $z_{\rm ion}$.
 Three lines are corresponding to the values of $z_{\rm ion}=7,8,$ and $9$,
 from down to up at $\ell=10$, respectively.
Middle-Right: $D^{BB}_\ell$ with different values of $\tau = 0.054 \pm 0.007$.
 Three lines are corresponding to those of the upper bound value, center value and lower bound value,
 from up to down, respectively.
Right: $D^{BB}_\ell$ with different value of $z_{\rm ion}$.
 Three lines are corresponding to the values of $z_{\rm ion}=7,8,$ and $9$,
 from down to up at $\ell=10$, respectively.
}
\label{fig:changin-tau-and-zion}
\end{figure}
Before closing this section
 we discuss the effect of ambiguity of our knowledge of reionization.
Fig.\ref{fig:changin-tau-and-zion}
 shows how angular power spectra change
 depending on the value of the optical depth $\tau$
 and the value of the redshift $z_{\rm ion}$ when reionization starts.
The ambiguity of the optical depth by the PLANCK collaboration
 causes the change of power spectra within cosmic variance.
The lack of the knowledge about reionization process gives larger distortions beyond cosmic variance,
 and it is important to obtain the knowledge by exploring deeper universe 
 through future observations of quasars, 21-cm hyperfine line of hydrogen, and so on.
Since we have to confront cosmic variance at low-$\ell$,
 the precise knowledge of reionization is necessary
 as well as more precise theoretical calculations of polarization power spectra.

\section{Conclusions}
\label{sec:conclusions}

There are many ways to measure the Hubble constant, the present expansion rate of the universe,
 and at present these results do not converge to a single preferable value within errors.
This problem so called Hubble tension may indicate something about dark energy of the universe
 which can change the way of expanding the universe.
We have concentrated on the possibility that
 some physics at lower redshift, around the period of reionization,
 may give a solution of the Hubble tension.
Many measurements of the Hubble parameter at various values of redshift
 suggest the history of expansion of the universe at lower redshift.
An important observation is that
 the measurements of the Hubble parameter at various redshifts
 can be categorized into two groups depending on how the distance is determined:
 by the standard distance ladder or the inverse distance ladder \cite{Cuesta:2014asa}. 
We have made a simple fit of the Hubble parameter as a function of redshift,
 by using only the data with the standard distance ladder
 or only the data with the inverse distance ladder.
In these fits we have introduced two phenomenological models
 of the equation of state of dark energy as a function of redshift:
 one is the CPL model and the other is the model of Taylor expansion.
Therefore,
 we have obtained four different data--driven phenomenological models of dark energy.

The polarizations of the CMB at low $\ell \lesssim 10$ are mainly produced
 through the Thomson scattering of the CMB photons off the free electrons
 which are produced during reionization.
The time evolution of the free electron density is determined
 by the reionization process as well as the way of expanding the universe
 in the period of the redshift $z \lesssim 10$.
Although our data--driven phenomenological models
 have been obtained by the data in the range of redshift $0 \lesssim z \lesssim 1$,
 we have assumed that they can be applied to the period of reionization by extrapolations.
We have developed the semi-analytic method to calculate the angular power spectrum
 of the E-mode polarization of the CMB due to scalar perturbations with long wave length limit,
 which is almost the same method in an author's previous work
 for the B-mode polarization due to tensor perturbations \cite{Kitazawa:2019fzc}.
The angular power spectra of E-mode and B-mode polarizations
 due to scalar and tensor perturbations, respectively,
 have been calculated in our four phenomenological models using this semi-analytic method. 
We have compared the results with the prediction of the concordance $\Lambda$CDM model.

The two models with the standard distance ladder
 give the angular power spectra which are somewhat enhanced at larger $\ell$.
On the other hand,
 the two models with the inverse distance ladder
 give the angular power spectra which are slightly suppressed at larger $\ell$.
There is no model dependence,
 whether the CPL model or the model of Taylor expansion, in this results,
 though their redshift dependences of the equation of state of dark energy
 are not exactly the same in the period of reionization.
This is because in that period the universe is in the era of matter dominant
 and the contribution of the difference is little.
The important difference for the result
 is whether $w$ is smaller or larger than $-1$ at smaller redshift,
 corresponding to the models with the standard distance ladder and the inverse distance ladder,
 respectively.
These arguments lead us to a model independent observation that
 in the production of the polarizations of the CMB in the period of reionization
 with non--trivially time--dependent equation of state of dark energy,
 the evolution of the universe suggested by the standard distance ladder indicates
 higher polarization powers at larger $\ell$,
 and the evolution of the universe suggested by the inverse distance ladder
 indicates lower polarization powers at larger $\ell$,
 than the prediction of the concordance $\Lambda$CDM model.
Therefore,
 future precise measurements of low-$\ell$ polarizations of the CMB by LiteBIRD, for example,
 may provide some indication to the Hubble tension.
In addition to the cosmic variance limited measurements
 it is also necessary to provide precise knowledge of the reionization process,
 the value of optical depth, the time to start reionization, and so on.
Even as of today
 more precise fits of Hubble parameter with more possible data
 in both with the standard and inverse distance ladders,
 and more precise numerical calculations of angular power spectra of CMB polarizations
 are worth to pursue this observation further.
Furthermore,
 discussion about practical measurements in future experiments (signal--to--noise ratio, for example)
 requires to include by numerical calculations other contributions to polarizations:
 E-mode polarizations which are produced at the period of recombination and
 B-mode polarizations which are produced by the lensing effect from E-mode polarizations.
We leave this analysis for future work.

\section*{Acknowledgments}

The author would like to thank A.~Gruppuso for helpful comments.
This work was supported in part by JSPS KAKENHI Grant Number 19K03851.

\end{document}